# Linear-Quadratic regulators for internal boundary control of lane-free automated vehicle traffic

Milad Malekzadeh[*], Ioannis Papamichail, Markos Papageorgiou

Dynamic Systems and Simulation Laboratory, Technical University of Crete, Chania 73100, Greece

[*]Corresponding author (email: mmalek@dssl.tuc.gr)

**Abstract**

Lane-free vehicle movement has been recently proposed for connected automated vehicles (CAV) due to various potential advantages. One such advantage stems from the fact that incremental changes of the road width in lane-free traffic lead to corresponding incremental changes of the traffic flow capacity. Based on this property, the concept of internal boundary control was recently introduced to flexibly share the total road width and capacity among the two traffic directions of a highway in real-time, in response to the prevailing traffic conditions, so as to maximize the cross-road (both directions) infrastructure utilization. Feedback-based Linear-Quadratic regulators with or without Integral action (LQI and LQ regulators) are appropriately developed in this paper to efficiently address the internal boundary control problem. Simulation investigations, involving a realistic highway stretch and different demand scenarios, demonstrate that the proposed simple regulators are robust and similarly efficient as an open-loop nonlinear constrained optimal control solution, while circumventing the need for accurate modelling and external demand prediction.

## 1. Introduction

Recurrent traffic congestion on urban freeways, highways and arterials is an increasingly serious problem for most big cities around the world, causing substantial delays, increased fuel consumption, excessive environmental pollution and reduced traffic safety. Traffic management measures, utilizing conventional means, are valuable (Papageorgiou et al., 2003; Kurzhanskiy and Varaiya, 2010) and, in some cases, able to delay or even avoid the onset of congestion. However, they are not always sufficient to tackle heavily congested traffic conditions. Gradually emerging and future ground-breaking vehicle automation and communication systems should be exploited to develop innovative solutions that can be applied within a smart road infrastructure.

Vehicle automation systems range from different kinds of driver support systems (e.g. ACC and lane-assist systems that are already commercially available with most car manufacturers) to highly or fully automated driving (i.e. SAE levels 4 and 5 vehicles); while vehicle communication enables vehicle-to-vehicle (V2V) and vehicle-to-infrastructure (V2I) communication that may support various potential applications. Most vehicle manufacturers and, lately, some information-technology companies (like Waymo/Google), as well as research institutions, have been developing and testing in real traffic conditions high-automation or virtually driverless autonomous vehicles that monitor their environment and make sensible driving decisions based on appropriate decision and control algorithms (Ardelt et al., 2012; Aeberhard et al., 2015; Kamal et al., 2016; Makantasis and Papageorgiou, 2018).

Recently, Papageorgiou et al. (2021) launched the TrafficFluid concept, a novel paradigm for vehicular traffic that is applicable at high penetration rates of vehicles equipped with high levels of vehicle automation and communication systems. The TrafficFluid concept suggests: (1) lane-free traffic, whereby vehicles are not bound to fixed traffic lanes, as in conventional traffic; (2) vehicle nudging, whereby vehicles may exert a "nudging" effect on, i.e. influence the movement of vehicles in front of them. Vehicles in a lane-free environment do not necessarily align to form lanes, but are self-organizing into dynamically changing 2-D clusters, depending on the vehicle sizes, their desired speeds, the employed vehicle movement strategies and the prevailing density, so as to maximize the available infrastructure utilization. Thus, if the road width is increased or decreased by some amount, the vehicles, driven by their movement strategy, spread accordingly to cover the changed 2-D road space, and new possible clusters may form. As demonstrated by Papageorgiou et al. (2021), if the road is widened by some amount, vehicles immediately move laterally to cover the free space. This increases the average inter-vehicle spacing, thus allowing for higher vehicle speeds and hence higher flow and capacity. In this context, the internal boundary control concept, introduced by Malekzadeh et al. (2021), exploits the lane-free principle of TrafficFluid, specifically the property that the road capacity may exhibit incremental (increasing or decreasing) changes in response to corresponding incremental (widening or narrowing) changes of the road width. This is in contrast to lane-based roads and traffic, where capacity changes may only occur if the road width is changed by a lane or a lane-multiple.

Consider a road with two opposite traffic directions serving connected automated vehicles (CAVs). The total available cross-road capacity (for both directions) may be shared between the two directions in a flexible way, according to the prevailing demand per direction, so as to maximize the infrastructure exploitation. Flexible capacity sharing may be achieved by virtually moving the internal boundary, which separates the two traffic directions, and communicating this decision to CAVs, so that they respect the changed road boundary. This way, the road width portion (and total capacity share) assigned to each direction can be changed in space and time (subject to constraints) according to an appropriate real-time control strategy, as illustrated in Figure 1, so as to maximize the total traffic efficiency of the overall system.

The idea of sharing the total cross-road capacity among the two traffic directions is not new and has been occasionally employed for conventional lane-based traffic, typically with manual interventions (see Wolshon and Lambert, 2006; Ampountolas et al. (2020) and references therein for a review of systems that exist around the globe). The measure is known as tidal flow (or contra-flow or reversible lanes) control, and its main principle is to adapt the total available cross-road supply to the demand per direction. Its most basic form is the steady allocation of one (or more) lanes of one direction to the other direction for a period of time in the aim of addressing abnormal traffic supply or demand in one traffic direction, e.g. at work zones or at big events, holiday departure or return, evacuation etc. More advanced reversible lane control systems may operate in real time, e.g. to balance delays on both sides of a known bottleneck (e.g. bridge, tunnel) by assigning a lane to one of the two directions in alternation in response to the prevailing traffic conditions. To this end, optimal control or feedback control algorithms of various types were proposed by Xue and Dong (2000), Frejo et al. (2015) and Ampountolas et al. (2020).

Reversible lanes have also been considered in connection with lane-based CAV driving. Duell et al. (2015) use the system optimal dynamic traffic assignment models formulated by Ziliaskopoulos (2000) for a single destination and by Li et al. (2003) for more general networks, using the Cell Transmission Model (CTM) by Daganzo (1994). Lanes are introduced as integer variables, and the problem is formulated as a mixed integer linear programming (MILP) problem that has, however, high (exponential) complexity due to the many integers variables involved. Levin and Boyles (2016) use this model for a single link and utilize stochastic demand as a Markov decision process. The MILP problem is solved using a heuristic and is incorporated within a UE routing problem.

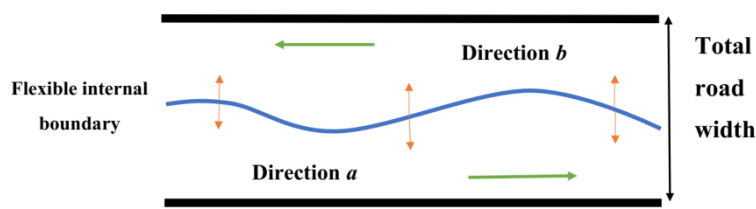

Figure 1: Space-time flexible internal road boundary.

The use of tidal flow control systems in lane-based traffic is not widespread for a number of reasons (see Malekzadeh et al. (2021)), such as: harsh resolution of infrastructure sharing (only by lane quanta) among the two traffic directions; serious counter-problems due to frequent merging or diverging traffic at lane-drop or lane-gain areas; safety-induced time-delays after each lane switch. These serious difficulties entail very limited capacity sharing flexibility in space and time and hinder reversible lane control from being a major traffic management measure. Even in the future CAV traffic, some of the mentioned difficulties would persist in lane-based conditions, notably the low capacity sharing resolution and the merging nuisance.

In contrast, in a lane-free CAV traffic environment, the mentioned difficulties are largely mitigated (Malekzadeh et al. (2021)). More specifically:

- The resolution of road-width sharing among the two directions can be high, still leading to corresponding intended capacity changes for the two opposite traffic directions.

- Assuming smooth CAV driving on a lane-free road surface, the internal boundary may be a smooth space-function, as illustrated in Figure 1, which may be smoothly changed in real time in response to the prevailing traffic conditions.

- Assuming moderate changes of the internal boundary over time and space, the aforementioned safety-induced time-delay may be very small.

Thanks to these characteristics, real-time internal boundary control for lane-free CAV traffic may be broadly applicable to the high number of arterial or highway infrastructures that feature unbalanced demands during the day in the two traffic directions, so as to strongly mitigate or even utterly avoid congestion. Even for infrastructures experiencing strong demand in both directions quasi-simultaneously, real-time internal boundary control may intensify the road utilization and lead to sensible improvements.

Malekzadeh et al. (2021) analyzed the internal boundary control problem and demonstrated its high improvement potential by formulating and solving a model-based open-loop nonlinear constrained optimal control problem, which requires accurate modelling and external demand prediction, in the form of a convex Quadratic Programming (QP) problem. That approach considers explicitly the optimization of a physical measure of performance Total Time Spent (TTS). Although the application of such a controller appears difficult in the field, it was developed in order to be able to know the upper limit of performance improvement that can be achieved for any scenario in a macroscopic simulation environment. In fact, the open-loop approach may be used within a Model

Predictive Control (MPC) frame with updated external demand predictions for real-time application; however, simpler real-time algorithms, that do not call for demand prediction and reach a similar efficiency level as the mentioned QP approach, are preferable. This paper develops and investigates the application of Linear-Quadratic regulators with or without Integral action (LQI and LQ regulators) for the internal boundary control problem. The well-known CTM is used, after linearization, for controller design; while its full nonlinear form is used for simulation testing of the developed controllers. Using the same highway stretch as in (Malekzadeh et al., 2021), a couple of well-designed and challenging demand scenarios are considered, and the performance of the LQ regulators is compared to each other, to the no-control case, as well as to the optimal results obtained using the QP formulation by Malekzadeh et al. (2021) with perfect prediction (upper limit of achievable performance improvement).

Section 2 presents some background issues, specifically the CTM equations and an outline of the QP problem formulation by Malekzadeh et al. (2021). Section 3 presents the design of the LQ and LQI regulators. Section 4 presents the simulation investigations, while conclusions are drawn in Section 5.

## 2. Background issues

### 2.1. Sharing factor

The internal boundary control problem should be designed in a macroscopic setting, so as to account for the traffic conditions on the highway and act accordingly. Lane-free traffic is not expected to give rise to structural changes of existing macroscopic traffic flow models. It is reasonable to assume, as also supported by results in (Bhavathrathan and Mallikarjuna, 2012; Asaithambi et al., 2016; Munigety et al., 2016; Papageorgiou et al., 2021), that notions and concepts like the conservation equation, the Fundamental Diagram (FD), as well as moving traffic waves will continue to characterize macroscopic traffic flow modelling in the case of CAV lane-free traffic. Additionally, specific physical traffic parameters, such as free speed, critical density, flow capacity, jam density, are also relevant for lane-free traffic, but may of course take different values than in lane-based traffic. The exact values that these parameters will take in CAV lane-free traffic are of minor importance for the presented demonstration of the novel internal boundary control measure.

Let us call the two opposite traffic directions, presented in Figure 1, directions $a$ and $b$, respectively. We assume that, at specific road sections, each direction is assigned a respective road width $w^a = \varepsilon \cdot w$ and $w^b = (1-\varepsilon) \cdot w$, where $0 \leq \varepsilon \leq 1$ is the sharing factor, to be specified in real time as a control input by the internal boundary controller, and $w$ is the total road width (both directions).

Let $Q(\rho)$, where $\rho$ is the traffic density in veh/km, be the FD of a road section, which would apply if the whole road width would be assigned to only one of the two opposite traffic directions (i.e. for $\varepsilon$ equal 0 or 1), with total critical density $\rho_{cr}$, total capacity $q_{cap}$ (in veh/h) and total jam density $\rho_{max}$. Let us now consider the case of partial road sharing, i.e. $\varepsilon_{min} \leq \varepsilon \leq \varepsilon_{max}$, where $\varepsilon_{min}, \varepsilon_{max} \in (0,1)$ are appropriate bounds aiming to suppress utter closure of either direction. As shown by Malekzadeh et al. (2021), the FDs for the two directions are functions of $\varepsilon$ given by

$$Q^a(\rho^a, \varepsilon) = \varepsilon \cdot Q(\rho^a / \varepsilon)$$
$$Q^b(\rho^b, \varepsilon) = (1-\varepsilon) \cdot Q(\rho^b / (1-\varepsilon))$$
(1)

where $\rho^a$ and $\rho^b$ (in veh/km) are the respective densities of the two directions. The critical density $\rho_{cr}^a(\varepsilon)$, capacity $q_{cap}^a(\varepsilon)$, and jam density $\rho_{max}^a(\varepsilon)$ for direction $a$ are thus functions of the sharing factor $\varepsilon$ and are given (Malekzadeh et al., 2021) by $\rho_{cr}^a(\varepsilon) = \varepsilon \cdot \rho_{cr}$, $q_{cap}^a(\varepsilon) = \varepsilon \cdot q_{cap}$ and $\rho_{max}^a(\varepsilon) = \varepsilon \cdot \rho_{max}$, i.e. the sharing factor $\varepsilon$ scales density (veh/km) and flow (veh/h), leaving unaffected the speed (km/h). The corresponding relations for direction $b$ are $\rho_{cr}^b(\varepsilon) = (1-\varepsilon) \cdot \rho_{cr}$, $q_{cap}^b(\varepsilon) = (1-\varepsilon) \cdot q_{cap}$ and $\rho_{max}^b(\varepsilon) = (1-\varepsilon) \cdot \rho_{max}$. All relations for the FD parameters hold for FDs that may be smooth or non-differentiable at $\rho_{cr}$, e.g. the triangular FD; see Figure 2 for illustration; see also (Ampountolas et al., 2020).

For controller design, a dynamic traffic flow model must be used. A simple but realistic option is CTM. CTM is a first-order dynamic traffic flow model with a triangular FD, which attains a space-time discretized form by application of the Godunov numerical scheme. The following section presents CTM and appropriate adjustments that have been introduced to incorporate capacity drop and the effect of the sharing factor $\varepsilon$.

## 2.2. Extended CTM

Consider a highway stretch with two reverse traffic directions $a$ (from left to right) and $b$ (from right to left). The stretch is subdivided into $n$ sections, with lengths $L_i$, $i = 1, 2, \ldots, n$. As explained in the previous section, the total road width, which is assumed constant over all sections for simplicity, can be flexibly shared among the two directions in real time. As the sharing may be different for every section, we have corresponding sharing factors $\varepsilon_i$, $i = 1, 2, \ldots, n$; and (1) applies to each section. As a consequence, the total section capacity, as well as the critical density and jam density, are shared among traffic directions $a$ and $b$ according to

$$q_{i,cap}^{a}(\varepsilon_i) = \varepsilon_i \cdot q_{cap}, \quad q_{i,cap}^{b}(\varepsilon_i) = (1-\varepsilon_i) \cdot q_{cap}$$
$$\rho_{i,cr}^{a}(\varepsilon_i) = \varepsilon_i \cdot \rho_{cr}, \quad \rho_{i,cr}^{b}(\varepsilon_i) = (1-\varepsilon_i) \cdot \rho_{cr} \qquad (2)$$
$$\rho_{i,\max}^{a}(\varepsilon_i) = \varepsilon_i \cdot \rho_{\max}, \quad \rho_{i,\max}^{b}(\varepsilon_i) = (1-\varepsilon_i) \cdot \rho_{\max}.$$

The corresponding changes of the triangular FD that may occur at each section and traffic direction are illustrated in Figure 2. More specifically, when the value of the sharing factor is 0.5, i.e., the flow capacities of the two directions are equal, their FDs are "nominal" (blue line with $(.)^N$ parameters); when the sharing factor is different than 0.5, we have two FDs: the extended one (green line with $(.)^E$ parameters) applies to the direction that is assigned more width and hence more flow capacity, and the reduced, complementary FD (orange line with $(.)^R$ parameters) applies to the other direction that is assigned less width and flow capacity. Based on (2), all FD parameters of a section change, whenever it is decided to change the corresponding sharing factor in real time. The exact values of the specific physical traffic parameters used in the CTM model, such as free speed, critical density, flow capacity, jam density, will depend on the vehicles' physical dimensions (width and length) and desired speed distribution, the total width of the road and the vehicle moving strategy employed for lane-free traffic.

As stated earlier, the sharing factors take values $\varepsilon_i \in [0,1]$. However, for the internal boundary control problem, we would like to disallow the utter closure of either direction; hence, the assigned road width in either direction should never be smaller than the widest vehicles driving on the road. This requirement gives rise to stricter constraints for the sharing factors as follows

$$0 < \varepsilon_{i,\min} \leq \varepsilon_i \leq \varepsilon_{i,\max} < 1 \qquad (3)$$

where $\varepsilon_{i,\min} \cdot w$ and $(1-\varepsilon_{i,\max}) \cdot w$ are the minimum admissible widths to be assigned to directions $a$ and $b$, respectively.

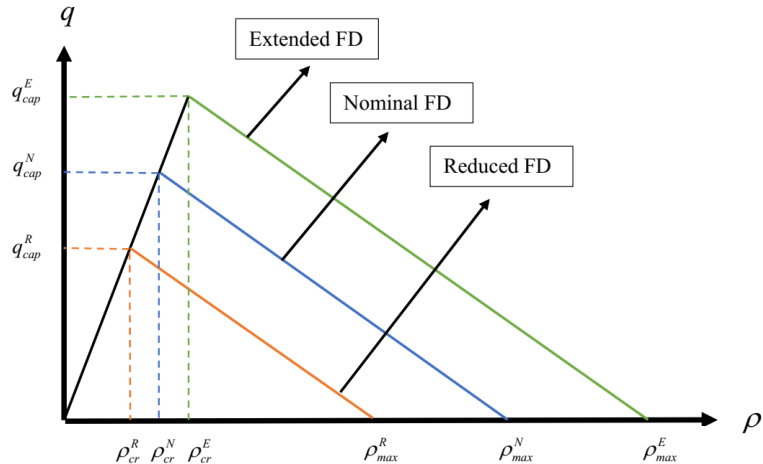

Figure 2: The triangular fundamental diagram with flexible internal boundary.

Another restriction to be applied to the sharing factors concerns the time-delay needed to evacuate traffic on the direction that receives a restricted width, compared with the previous control time-step. As discussed earlier, this time-delay is small in lane-free CAV traffic without physical barrier among the two traffic directions and with moderate changes of the sharing factors applied to short sections, but needs nevertheless to be considered. Clearly, the time-delay should apply only to the traffic direction that is being widened, compared to the previous control interval; while the direction that is restricted should promptly apply the smaller width, so that CAVs therein move out of the reduced-width zone. Assume that the required time-delay is smaller than or equal to the control time interval $T_c$; then, the time-delay requirement is automatically fulfilled for each section $i$, if the sharing factors that are actually applied to the two directions, i.e. $\varepsilon_i^a$ and $\varepsilon_i^b$, respectively, are calculated as follows

$$\varepsilon_i^a(k_c) = \min\{\varepsilon_i(k_c), \varepsilon_i(k_c - 1)\} \tag{4}$$

$$\varepsilon_i^b(k_c) = \min\{1 - \varepsilon_i(k_c), 1 - \varepsilon_i(k_c - 1)\} \tag{5}$$

where $k_c = 0, 1, \ldots$ is the discrete control time index. It is noted that the notation $\varepsilon_i^a(k_c)$ and $\varepsilon_i^b(k_c)$ indicates that the sharing factors are applied for the duration of the control time interval $[k_c \cdot T_c, (k_c + 1) \cdot T_c)$. The above equations may be readily extended if the required time-delay is a multiple of the control time interval $T_c$.

Traffic flows from section 1 to section $n$ in direction $a$; and from section $n$ to section 1 in direction $b$ (see Figure 3 as an example). We denote $\rho_i^a$, $i = 1, 2, \ldots, n$, the traffic density of section $i$, direction $a$; and $\rho_i^b$, $i = 1, 2, \ldots, n$, the traffic density of section $i$, direction $b$. Similarly, we denote $q_i^a$, $i = 1, 2, \ldots, n$, and $q_i^b$, $i = 1, 2, \ldots, n$, the mainstream exit flows of section $i$ for directions $a$ and $b$, respectively. Thus, $q_0^a$ is the feeding upstream mainstream inflow for direction $a$; and $q_{n+1}^b$ is the feeding upstream mainstream inflow for direction $b$. Every section, except for the most upstream in each direction, may have an on-ramp or an off-ramp at its upstream boundary. The on-ramp flows (if any) at section $i$ are denoted $r_i^a$ for direction $a$, and $r_i^b$ for direction $b$. The off-ramp flow (if any) of section $i$, direction $a$, is calculated based on known exit rates $\beta_i^a$ multiplied with the upstream-section flow, i.e. $\beta_i^a q_{i-1}^a$; and the off-ramp flow (if any) of section $i$, direction $b$, is calculated based on known exit rates $\beta_i^b$ multiplied with the upstream-section flow, i.e. $\beta_i^b q_{i+1}^b$.

The conservation equations for the sections of direction $a$ are:

$$\rho_1^a(k+1) = \rho_1^a(k) + \frac{T}{L_i}(q_0^a(k) - q_1^a(k))$$

$$\rho_i^a(k+1) = \rho_i^a(k) + \frac{T}{L_i}((1-\beta_i^a)q_{i-1}^a(k) - q_i^a(k) + r_i^a(k)), \; i = 2, 3, \ldots, n \quad (6)$$

where $T$ is the model time-step, typically set equal to $5 - 10$ s for section lengths of some 500 m in length, and $k = 0, 1, \ldots$ is the corresponding discrete time index of the model.

According to CTM, traffic flow is obtained as the minimum of demand and supply functions, except for the last section, where only the demand function is considered, assuming that the downstream traffic conditions are uncongested. Clearly, when writing the demand and supply functions $Q_D$ and $Q_S$, respectively, for the case of the internal boundary control problem, we need to consider the impact of the respective sharing factors $\varepsilon_i^a(k_c)$ on the FDs. Thus we have

$$q_i^a(k) = \min\left\{Q_D(\rho_i^a(k), \varepsilon_i^a(k_c)), \frac{Q_S(\rho_{i+1}^a(k), \varepsilon_{i+1}^a(k_c))}{(1-\beta_{i+1}^a)} - \lambda_r r_{i+1}^a(k)\right\}, \; i = 1, 2, \ldots, n-1$$

$$q_n^a(k) = Q_D(\rho_n^a(k), \varepsilon_i^a(k_c)). \quad (7)$$

The demand and supply functions are given by the following respective equations

$$Q_D(\rho, \varepsilon) = \min\left\{\varepsilon q_{cap} + \lambda_d q_{cap} \frac{\rho - \varepsilon \rho_{cr}}{\rho_{cr} - \rho_{max}}, v_f \rho\right\},$$

$$Q_S(\rho, \varepsilon) = \min\left\{\varepsilon q_{cap}, w_s(\varepsilon \rho_{max} - \rho)\right\}, \quad (8)$$

where $v_f$ is the free speed (which is assumed equal for all sections for simplicity) and $w_s$ is the back-wave speed. The control time-step $T_c$ does not need to be equal to the model time-step, but is assumed to be a multiple of $T$, in which case the control time index is given by $k_c = \lfloor kT/T_c \rfloor$, where $\lfloor . \rfloor$ is the integer part notation.

It is well-known that CTM does not reproduce the capacity drop phenomenon, i.e. the empirical finding that, at the head of congestion, the observed flow in real traffic is reduced compared to the road capacity. Capacity drop is deemed to occur in conventional traffic due to bounded and differing accelerations of different vehicles (Yuan et al., 2015). Recently, CTM has been extended in a number of possible ways to enable the reproduction of capacity drop, see (Kontorinaki et al., 2017) for an overview and comparison. The presence of capacity drop in conventional freeway traffic is a major reason for infrastructure degradation and for the need of introducing traffic control measures to restore capacity (Papageorgiou et al., 2003). In contrast, in the present context of internal boundary control, the presence of capacity drop marks a secondary source of amelioration of the traffic conditions, because the potential benefits achievable via opportune capacity sharing are expected to be much higher than those resulting from capacity drop avoidance. In fact, it is unknown at the moment, if and to what extent capacity drop may occur in lane-free CAV traffic. To be able to investigate the impact of possible capacity drop, we have incorporated in the above equations the option of introducing capacity drop trough appropriate terms according to Kontorinaki et al. (2017). More specifically, this option is enabled via the parameters $\lambda_r$ and $\lambda_d$ in equations (7) and (8). If these parameters are set to $\lambda_r = 1$ and $\lambda_d = 0$, no capacity drop is reproduced, as typical for CTM; if these values are set between 0 and 1, a corresponding level of capacity drop is produced by the model.

The equations for direction $b$ are analogous to those of direction $a$, with few necessary index modifications. Section numbers in direction $b$ are descending, hence we have

$$\rho_i^b(k+1) = \rho_i^b(k) + \frac{T}{L_i}((1-\beta_i^b)q_{i+1}^b(k) - q_i^b(k) + r_i^b(k)), i = 1, 2, \ldots, n-1$$

$$\rho_n^b(k+1) = \rho_n^b(k) + \frac{T}{L_i}(q_{n+1}^b(k) - q_n^b(k))$$

(9)

and the flows are given by

$$q_1^b(k) = Q_D(\rho_1^b(k), \varepsilon_i^b(k_c))$$

$$q_i^b(k) = \min\left\{Q_D(\rho_i^b(k), \varepsilon_i^b(k_c)), \frac{Q_S(\rho_{i-1}^b(k), \varepsilon_{i-1}^b(k_c))}{(1-\beta_{i-1}^b)} - \lambda_r r_{i-1}^b(k)\right\}, i = 2, 3, \ldots, n$$

(10)

## 2.3. The QP problem formulation

To be able to assess the level of performance of the LQ regulators to be developed, we will compare its results with those obtained with a nonlinear constrained optimal control scheme developed for the same problem by Malekzadeh et al. (2021). That scheme is based on the formulation and solution of a convex Quadratic Programming (QP) problem that is outlined here for completeness.

The CTM, presented in Section 2.2, is utilized in the problem formulation. Due to the presence of the min-operators in (7), (8) and (10), CTM is a nonlinear model. As proposed by Papageorgiou (1995) and practiced in most previous utilizations of CTM for optimal control (Ziliaskopoulos, 2000; Gomes and Horowitz, 2006; Roncoli et al., 2015), such nonlinearities may be transformed to linear inequalities by requesting the left-hand side of the equation, where the min-operator appears, to be smaller than or equal to each one of the terms included in the min-operator. As a result, the above mentioned equations of CTM are replaced in this formulation by respective groups of linear inequalities for each section and direction. Similarly, each one of the equations (4) and (5) is replaced by two linear inequalities. Finally, the constraints (3) must be considered at each discrete point to appropriately limit the sharing factors.

All these inequalities, presented in detail by Malekzadeh et al. (2021), transform the CTM equations, including the changing sharing factors and their constraints, to a set of linear equalities and inequalities, which thus constitute a convex feasible region for an optimization problem. Combining this model with a convex quadratic objective function, to be minimized, results in a QP formulation that can be solved numerically using very efficient available codes. The objective function is as follows:

$$J_{QP} = T\sum_{k=1}^{K}\sum_{i=1}^{n}\left(L_i\rho_i^a(k) + L_i\rho_i^b(k)\right) - w_1\sum_{k_c=0}^{K_c-1}\sum_{i=1}^{n}\left(\varepsilon_i^a(k_c) + \varepsilon_i^b(k_c)\right)$$
$$+ w_2\sum_{k_c=1}^{K_c-1}\sum_{i=1}^{n}\left(\varepsilon_i(k_c) - \varepsilon_i(k_c-1)\right)^2 + w_3\sum_{k_c=0}^{K_c-1}\sum_{i=2}^{n}\left(\varepsilon_i(k_c) - \varepsilon_{i-1}(k_c)\right)^2 \quad (11)$$
$$+ w_4\sum_{k_c=0}^{K_c-1}\sum_{i=1}^{n}\left(\frac{\varepsilon_i(k_c)^2}{\hat{d}_i^a(k_c)} + \frac{(1-\varepsilon_i(k_c))^2}{\hat{d}_i^b(k_c)}\right).$$

The cost function extends over a time horizon of $K$ model time-steps or $K_c$ control time-steps; it includes five terms, the first two being linear and the rest of them quadratic. The first term is the most important one and represents the Total Time Spent (TTS) that determines the traffic efficiency resulting from the proposed control actions. The second term ensures that one of the two inequalities introduced to replace equation (4), and one of the two inequalities introduced to replace equation (5) will be activated and, as a result, one of the two terms included in the min-operators of equations (4) and (5) will actually materialize. The following three terms are quadratic and reflect secondary operational and policy objectives. The first of these quadratic terms penalizes the variation of the control input in consecutive time-steps, so that changes of the internal boundary of each section from one control time-step to the next remain small. The second quadratic term penalizes the space variation of the control input from section to section, so as to mitigate strong changes in the road width assigned to each direction within a short distance. The last quadratic term is policy related and attempts to assign to the two directions respective capacity shares that balance the respective capacity reserves for each section. To this end, we use $\hat{d}_i^a(k_c)$ and $\hat{d}_i^b(k_c)$, $i=1,..,n$, which are the projected demand trajectories for each section of the two respective directions. The weights $w_1, w_2, w_3, w_4$ were tuned using various demand scenarios in order to guarantee that the minimum achievable TTS value is always achieved. More details are provided by Malekzadeh et al. (2021).

## 3. Design of the LQ regulators

### 3.1. Relative densities

In conventional traffic management, traffic densities have a central role, as they reflect unequivocally the state of traffic, e.g. on the FD. In the novel internal boundary control setting, however, the traffic density variables $\rho_i^a$ and $\rho_i^b$ (in veh/km) in the two opposite directions of each section $i$ are not directly indicating the traffic conditions (e.g. under-critical or congested) encountered. This is because the critical density for each direction is a function of the sharing factor and is changing according to the applied control action; thus, the same density value may mark an under-critical or congested traffic state, depending on the value of the critical density; or, equivalently, depending on the share of road width among the two directions.

Therefore, we proceed with the definition of new variables that are appropriate to reflect the actual traffic conditions in the internal boundary control context. The new variables are the *relative densities* (dimensionless) and are defined per section and per direction as the ordinary densities. The relative density of section $i$ and direction $a$ or $b$ is obtained by dividing the corresponding traffic density with the corresponding critical density, which, on its turn, depends on the sharing factor prevailing during the last time-step. Considering (2), we get the following relations for the relative densities in the two directions

$$\tilde{\rho}_i^a(k) = \frac{\rho_i^a(k)}{\varepsilon_i(k-1)\rho_{cr}}, \quad \tilde{\rho}_i^b(k) = \frac{\rho_i^b(k)}{(1-\varepsilon_i(k-1))\rho_{cr}}, \quad i=1,2,\ldots,n. \tag{12}$$

Note that, according to the respective variable definitions, which are typical in discrete-time modelling, the sharing factor $\varepsilon_i(k-1)$ applies during the time interval $[(k-1)T, kT)$; while a density $\rho_i^a(k)$ or $\rho_i^b(k)$ is an instantaneous value at time instant $kT$; and these definitions explain the choice of the discrete-time arguments in (12).

As evidenced with equations (4) and (5), a control time delay must be applied only to the direction that is being widened, not to the direction that is being narrowed. However, this distinction is not possible in the design of the LQ regulators, as (4) and (5) cannot be explicitly considered in the linearized control design model. Thus, we have the option of either applying the time delay to both traffic directions or to none. Both options were tried and resulted in similar results. Therefore, given that the first option complicates the design model, we opted for the second option in (12). Please note that the correct control time delays are always applied, according to equations (4) and (5), in the simulation model while testing the regulators.

The relative densities reflect unequivocally the state of traffic in the internal boundary control context. Specifically, the relative density of a section and direction reflects under-critical traffic conditions if it is in the range $[0,1)$; capacity flow if it equals 1; and over-critical traffic conditions if it is bigger than 1.

**3.2. Linearized model**

In order to derive the LQ regulators, we need to include in our problem a discrete-time linearized system. To this end, we use as a basis the CTM equations delivered in the previous section. Replacing (12) in (6), i.e. in the conservation equation for direction $a$, we get

$$\varepsilon_i(k)\rho_{cr}\tilde{\rho}_i^a(k+1) = \varepsilon_i(k-1)\rho_{cr}\tilde{\rho}_i^a(k) + \frac{T}{L_i}((1-\beta_i^a)q_{i-1}^a(k) - q_i^a(k) + r_i^a(k)), i = 2,3,\ldots,n$$

$$\varepsilon_1(k)\rho_{cr}\tilde{\rho}_1^a(k+1) = \varepsilon_1(k-1)\rho_{cr}\tilde{\rho}_1^a(k) + \frac{T}{L_i}(q_0^a(k) - q_1^a(k))$$

(13)

and by dividing with $\varepsilon_i(k-1)\rho_{cr}$ we get

$$\frac{\varepsilon_i(k)}{\varepsilon_i(k-1)}\tilde{\rho}_i^a(k+1) = \tilde{\rho}_i^a(k) + \frac{T}{L_i\rho_{cr}}\left((1-\beta_i^a)\frac{q_{i-1}^a(k)}{\varepsilon_i(k-1)} - \frac{q_i^a(k)}{\varepsilon_i(k-1)} + \frac{r_i^a(k)}{\varepsilon_i(k-1)}\right), i = 2,3,\ldots,n$$

$$\frac{\varepsilon_1(k)}{\varepsilon_1(k-1)}\tilde{\rho}_1^a(k+1) = \tilde{\rho}_1^a(k) + \frac{T}{L_i\rho_{cr}}\left(\frac{q_0^a(k)}{\varepsilon_1(k-1)} - \frac{q_1^a(k)}{\varepsilon_1(k-1)}\right).$$

(14)

It is now assumed that the values of the control inputs do not change significantly in consecutive sample times. In fact, as we will see at the end of this section, the control time-step is a multiple $M$ of the model time-step, hence the control input changes only every $M$ model time-steps. In addition, due to the quadratic penalization of the control inputs within the objective function (29), that is defined later, the control input changes moderately over time (at each control interval). As a result, it is reasonable to assume that $\varepsilon_i(k)/\varepsilon_i(k-1) \simeq 1$, $i = 1,2,\ldots,n$. This assumption simplifies significantly the equations and decreases the burden of derivative calculations necessary for the linearization of the model, without really affecting the control results achieved, as found in corresponding result comparisons. Defining also the one-step retarded control input as a new state variable according to

$$\gamma_i(k+1) = \varepsilon_i(k), i = 1,2,\ldots,n,$$

(15)

we finally get the state equations that replace the conservation equations for direction $a$

$$\tilde{\rho}_i^a(k+1) = \tilde{\rho}_i^a(k) + \frac{T}{L_i \rho_{cr}} \left( (1-\beta_i^a) \frac{q_{i-1}^a(k)}{\gamma_i(k)} - \frac{q_i^a(k)}{\gamma_i(k)} + \frac{r_i^a(k)}{\gamma_i(k)} \right), i = 2, 3, \ldots, n$$
$$\tilde{\rho}_1^a(k+1) = \tilde{\rho}_1^a(k) + \frac{T}{L_1 \rho_{cr}} \left( \frac{q_0^a(k)}{\gamma_1(k)} - \frac{q_1^a(k)}{\gamma_1(k)} \right).$$
(16)

Similarly, replacing (12) in (9), i.e. in the conservation equation for direction $b$, and making the same assumption for the value of the control input in consecutive sample times, we get the state equations that replace the conservation equations for direction $b$

$$\tilde{\rho}_i^b(k+1) = \tilde{\rho}_i^b(k) + \frac{T}{L_i \rho_{cr}} \left( (1-\beta_i^b) \frac{q_{i+1}^b(k)}{1-\gamma_i(k)} - \frac{q_i^b(k)}{1-\gamma_i(k)} + \frac{r_i^b(k)}{1-\gamma_i(k)} \right), i = 1, 2, \ldots, n-1$$
$$\tilde{\rho}_n^b(k+1) = \tilde{\rho}_n^b(k) + \frac{T}{L_n \rho_{cr}} \left( \frac{q_{n+1}^b(k)}{1-\gamma_n(k)} - \frac{q_n^b(k)}{1-\gamma_n(k)} \right).$$
(17)

According to CTM, traffic flow is obtained as the minimum of the demand and supply functions. Starting from CTM, we need to come up with an approximate linearized system to allow for the derivation of feedback controllers by use of the LQ regulator methodology. This is a procedure with a long history in traffic management and control, see e.g. (Isaksen and Payne, 1973; Papageorgiou, 1984; Papageorgiou et al., 1990; Diakaki and Papageorgiou, 1994; Aboudolas and Geroliminis, 2013), to name just a few. In such works, the nominal linearization state (density) is taken to be uncongested and close to the critical density. Following this procedure in our case, we assume that traffic flow is nominally operating around capacity and is determined only by the demand function; which however is also obtained as the minimum of two terms according to (8), namely as the minimum of the flow capacity (without the capacity-drop term) and the FD flow-density relation. In conventional traffic control, the flow capacity is constant; hence a linear or linearized flow-density relation is taken from the FD to characterize the linearized flow dynamics. In contrast, in the internal boundary control case, the capacity is directly proportional to the control input (sharing factor), as evidenced from (8), hence this term is deemed more significant in the linearized approximation of the system dynamics.

To confirm these arguments, we have considered, in preliminary simulation testing, three cases: (i) Use of the FD flow-density relation only; (ii) Use of the capacity term only (omitting the capacity drop term); and (iii) Use of a convex combination of the two terms included in the min-operator of the demand function. The investigation confirmed that case (iii) is indeed most efficient and robust for all traffic conditions.

In conclusion, to proceed with the linearized approximation of system dynamics, $q_i^a(k)$ and $q_i^b(k)$ are given by

$$q_i^a(k) = \sigma \varepsilon_i(k) q_{cap} + (1-\sigma) v_f \rho_i^a(k)$$
$$q_i^b(k) = \sigma(1-\varepsilon_i(k)) q_{cap} + (1-\sigma) v_f \rho_i^b(k)$$
(18)

where $0 \leq \sigma \leq 1$ is the convex combination parameter used. Taking into account (12) and (15), (18) can be rewritten as

$$q_i^a(k) = \sigma \varepsilon_i(k) q_{cap} + (1-\sigma) v_f \tilde{\rho}_i^a(k) \gamma_i(k) \rho_{cr}$$
$$q_i^b(k) = \sigma(1-\varepsilon_i(k)) q_{cap} + (1-\sigma) v_f \tilde{\rho}_i^b(k)(1-\gamma_i(k)) \rho_{cr}$$
(19)

Considering these equations, the system of dynamic equations (15)-(17) can be rewritten in the following state-space form:

$$\tilde{\rho}_i^a(k+1) = f_i^a\left(\tilde{\rho}_i^a(k), \tilde{\rho}_{i-1}^a(k), \varepsilon_i(k), \varepsilon_{i-1}(k), \gamma_i(k), \gamma_{i-1}(k), d_i^a(k)\right), i = 2,3,\ldots,n$$
$$\tilde{\rho}_1^a(k+1) = f_1^a\left(\tilde{\rho}_1^a(k), \varepsilon_1(k), \gamma_1(k), d_1^a(k)\right)$$
(20)

$$\tilde{\rho}_i^b(k+1) = f_i^b\left(\tilde{\rho}_i^b(k), \tilde{\rho}_{i+1}^b(k), \varepsilon_i(k), \varepsilon_{i+1}(k), \gamma_i(k), \gamma_{i+1}(k), d_i^b(k)\right), i = 1,2,\ldots,n-1$$
$$\tilde{\rho}_n^b(k+1) = f_n^b\left(\tilde{\rho}_n^b(k), \varepsilon_n(k), \gamma_n(k), d_n^b(k)\right)$$
(21)

$$\gamma_i(k+1) = g_i(\varepsilon_i(k)), i = 1,\ldots,n$$
(22)

where the upstream mainstream inflow, as well as the on-ramp flows of each direction, which are external inputs (disturbances), have been included in the respective demand vectors $\mathbf{d}^a(k) = \left[q_0^a(k), r_2^a(k), \ldots, r_n^a(k)\right]^T$ and $\mathbf{d}^b(k) = \left[r_1^b(k), \ldots, r_{n-1}^b(k), q_{n+1}^b(k)\right]^T$.

Linearization of the system (20)-(22) around a nominal point, denoted by superscript $N$, yields

$$\Delta \tilde{\rho}_i^a(k+1) = \left.\frac{\partial f_i^a}{\partial \tilde{\rho}_i^a}\right|_N \Delta \tilde{\rho}_i^a(k) + \left.\frac{\partial f_i^a}{\partial \tilde{\rho}_{i-1}^a}\right|_N \Delta \tilde{\rho}_{i-1}^a(k) + \left.\frac{\partial f_i^a}{\partial \varepsilon_i}\right|_N \Delta \varepsilon_i(k) + \left.\frac{\partial f_i^a}{\partial \varepsilon_{i-1}}\right|_N \Delta \varepsilon_{i-1}(k)$$
$$+ \left.\frac{\partial f_i^a}{\partial \gamma_i}\right|_N \Delta \gamma_i(k) + \left.\frac{\partial f_i^a}{\partial \gamma_{i-1}}\right|_N \Delta \gamma_{i-1}(k), i = 2,3,\ldots,n$$
(23)

$$\Delta \tilde{\rho}_1^a(k+1) = \left.\frac{\partial f_1^a}{\partial \tilde{\rho}_1^a}\right|_N \Delta \tilde{\rho}_1^a(k) + \left.\frac{\partial f_1^a}{\partial \varepsilon_1}\right|_N \Delta \varepsilon_1(k) + \left.\frac{\partial f_1^a}{\partial \gamma_1}\right|_N \Delta \gamma_1(k)$$

$$\Delta \tilde{\rho}_i^b(k+1) = \left.\frac{\partial f_i^b}{\partial \tilde{\rho}_i^b}\right|_N \Delta \tilde{\rho}_i^b(k) + \left.\frac{\partial f_i^b}{\partial \tilde{\rho}_{i+1}^b}\right|_N \Delta \tilde{\rho}_{i+1}^b(k) + \left.\frac{\partial f_i^b}{\partial \varepsilon_i}\right|_N \Delta \varepsilon_i(k) + \left.\frac{\partial f_i^b}{\partial \varepsilon_{i+1}}\right|_N \Delta \varepsilon_{i+1}(k)$$
$$+ \left.\frac{\partial f_i^a}{\partial \gamma_i}\right|_N \Delta \gamma_i(k) + \left.\frac{\partial f_i^a}{\partial \gamma_{i+1}}\right|_N \Delta \gamma_{i+1}(k), i = 1,2,\ldots,n-1$$
(24)

$$\Delta \tilde{\rho}_n^b(k+1) = \left.\frac{\partial f_n^b}{\partial \tilde{\rho}_n^b}\right|_N \Delta \tilde{\rho}_n^b(k) + \left.\frac{\partial f_n^b}{\partial \varepsilon_n}\right|_N \Delta \varepsilon_n(k) + \left.\frac{\partial f_n^a}{\partial \gamma_n}\right|_N \Delta \gamma_n(k)$$

$$\Delta \gamma_i(k+1) = \left.\frac{\partial g_i}{\partial \varepsilon_i}\right|_N \Delta \varepsilon_i(k), i = 1,2,\ldots,n$$
(25)

where $\Delta(.)(k) = (.)(k) - (.)^N$. Additionally, it has been assumed that $\Delta(.)(k) = 0$ for all disturbances, i.e. for the elements of the demands $\mathbf{d}^a(k)$ and $\mathbf{d}^b(k)$. The derivation of the partial derivatives necessary in (23)-(25), is presented in Appendix A.

Merging all the above, we obtain a linear state-space model

$$\Delta\mathbf{x}(k+1) = \mathbf{A}\Delta\mathbf{x}(k) + \mathbf{B}\Delta\mathbf{u}(k) \tag{26}$$

where $\Delta\mathbf{x}(k) = [\Delta\tilde{\boldsymbol{\rho}}^a(k)^T, \Delta\tilde{\boldsymbol{\rho}}^b(k)^T, \Delta\boldsymbol{\gamma}(k)^T]^T$ is the state vector and $\Delta\mathbf{u}(k) = \Delta\boldsymbol{\varepsilon}(k)$ is the control input vector, whereby $\Delta\tilde{\boldsymbol{\rho}}^a(k) = [\Delta\tilde{\rho}_1^a(k), \ldots, \Delta\tilde{\rho}_n^a(k)]^T$, $\Delta\tilde{\boldsymbol{\rho}}^b(k) = [\Delta\tilde{\rho}_1^b(k), \ldots, \Delta\tilde{\rho}_n^b(k)]^T$, $\Delta\boldsymbol{\gamma}(k) = [\Delta\gamma_1(k), \ldots, \Delta\gamma_n(k)]^T$ and $\Delta\boldsymbol{\varepsilon}(k) = [\Delta\varepsilon_1(k), \ldots, \Delta\varepsilon_n(k)]^T$. $\mathbf{A} \in \mathbb{R}^{3n \times 3n}$ and $\mathbf{B} \in \mathbb{R}^{3n \times n}$ are the state and input matrices, respectively.

If the control time-step is defined as a multiple of the model time-step, i.e. $T_c = MT$, where $M$ is an integer, then we have $k_c = \lfloor kT/T_c \rfloor = \lfloor k/M \rfloor$. Thus, the linear state-space equation may be changed as follows, in order to be based on the control time-step $T_c$,

$$\Delta\mathbf{x}(k_c+1) = \mathbf{A}^M \Delta\mathbf{x}(k_c) + \underbrace{(\mathbf{A}^{M-1} + \mathbf{A}^{M-2} + \ldots + \mathbf{I})\mathbf{B}}_{\hat{\mathbf{B}}} \Delta\mathbf{u}(k_c). \tag{27}$$

### 3.3. Quadratic cost function and integration states

The QP problem formulation of Section 2.3 provides sufficient freedom for considering, in the cost objective, the main physical traffic efficiency metric (TTS), as well as several secondary operational sub-objectives. In contrast, when employing the LQ methodology, the objective function is restricted to be a weighted quadratic norm of the states and of the control inputs. Nevertheless, we wish to achieve with the designed LQ regulators similarly efficient control results in terms of TTS, although TTS is not an explicit formal objective in the LQ problem formulation.

The state vector in (27) includes as state variables $\Delta\tilde{\rho}_i^a(k)$ and $\Delta\tilde{\rho}_i^b(k)$. By setting the nominal value of relative densities equal to 1, the squares of those state variables, to be included in the quadratic objective function, will motivate the resulting controller to operate the system near capacity, which is good for traffic efficiency. In addition, if capacity flow is not feasible (due to lack of demand or due to excessive demand), then minimizing a sum of squares has the tendency to balance deviations from the nominal values, something that is conform with the secondary operational sub-objective of balancing the margin to capacity across sections.

The quadratic objective also includes the sum of squares of the control inputs, i.e. of $\Delta\varepsilon_i(k)$. Using 0.5 as a nominal value for the sharing factors, minimization will have two effects: (i) to mitigate the deviations of the sharing factors from 0.5; and (ii) to balance these deviations in space (sections) and time, which are also secondary operational sub-objectives.

A final specification concerns the possible inclusion of integral terms in the regulator (LQI regulator), as often practiced in traffic control, see (Papageorgiou et al., 1990; Diakaki and Papageorgiou, 1994, Aboudolas and Geroliminis, 2013). The LQI regulator offers the possibility to drive to zero, in the steady state, a number of linear combinations of state variables, which cannot be higher than the number of control variables. In our case, the number of control variables equals the number of highway sections, and a good choice for the linear combinations to be considered could be $\Delta \tilde{\rho}_i^a(k) - \Delta \tilde{\rho}_i^b(k)$, which equals $\tilde{\rho}_i^a(k) - \tilde{\rho}_i^b(k)$ for nominal state values equal to 1. Striving for the same value of the relative densities on the two directions at each section seems equitable, and also seems to foster further the aforementioned balancing of capacity margins.

In view of the above, we may now proceed with the formal design of an LQI regulator. To enable the inclusion of integral terms in the regulator, we consider the state equation (27) augmented by use of the integrators

$$\mathbf{y}(k_c+1) = \mathbf{y}(k_c) + \underbrace{[\mathbf{I}_n, -\mathbf{I}_n, \mathbf{0}_{n \times n}]}_{\mathbf{H}} \underbrace{\begin{bmatrix} \Delta \tilde{\boldsymbol{\rho}}^{\mathbf{a}}(k_c) \\ \Delta \tilde{\boldsymbol{\rho}}^{\mathbf{b}}(k_c) \\ \Delta \boldsymbol{\gamma}(k_c) \end{bmatrix}}_{\Delta \mathbf{x}(k_c)}. \qquad (28)$$

We recall that, with this choice of integration state variables, we achieve a steady-state error equal to zero for the differences between the relative densities per direction for each section.

The control goal is to minimize the quadratic criterion

$$J_{LQI} = \frac{1}{2} \sum_{k_c=0}^{\infty} \left[ \|\Delta \mathbf{x}(k_c)\|_{\mathbf{Q}}^2 + \|\mathbf{y}(k_c)\|_{\mathbf{S}}^2 + \|\Delta \mathbf{u}(k_c)\|_{\mathbf{R}}^2 \right] \qquad (29)$$

where $\mathbf{Q} \in \mathbb{R}^{3n \times 3n}$, $\mathbf{S} \in \mathbb{R}^{n \times n}$, $\mathbf{R} \in \mathbb{R}^{n \times n}$ are symmetric positive definite matrices, which will be chosen in Section 4 to be diagonal for simplicity of tuning. The first term penalizes deviations of the elements of the state variable $\Delta \mathbf{x}$ from zero, i.e. deviations of $\tilde{\rho}_i^a(k)$, $\tilde{\rho}_i^b(k)$, $\gamma_i(k)$, $i = 1, 2, \ldots, n$, from their desired nominal values. The second term penalizes deviations of the elements of the state variable $\mathbf{y}$ from zero, i.e. deviations of the integral of the differences between the relative densities per direction for each section from zero. Finally, the third term penalizes deviations of the control inputs from the nominal values (i.e. from 0.5).

Considering the discrete-time linear system (27), (28) and the quadratic criterion (29), the following augmented problem matrices are defined

$$\tilde{\mathbf{A}} = \begin{bmatrix} \mathbf{A}^M & \mathbf{0}_{3n \times n} \\ \mathbf{H} & \mathbf{I}_n \end{bmatrix}, \tilde{\mathbf{B}} = \begin{bmatrix} \hat{\mathbf{B}} \\ \mathbf{0}_{n \times n} \end{bmatrix}, \tilde{\mathbf{Q}} = \begin{bmatrix} \mathbf{Q} & \mathbf{0}_{3n \times n} \\ \mathbf{0}_{n \times 3n} & \mathbf{S} \end{bmatrix}, \tilde{\mathbf{R}} = \mathbf{R}. \quad (30)$$

It is noted that, if the weighting matrix $\mathbf{S}$ is set equal to a zero matrix, then the solution of the above problem with state $\Delta\mathbf{x}$ and control $\Delta\mathbf{u}$ yields the solution of the LQ problem without integration terms, i.e. an LQ regulator (without integral parts).

### 3.4. Derivation of the LQ regulators

It can be shown that, for a convex combination parameter $0 \leq \sigma < 1$ used in (19), the linear system used is controllable. Given the above specifications, the optimal controller minimizing the criterion is given by

$$\Delta\mathbf{u}(k_c) = -\mathbf{K} \begin{bmatrix} \Delta\mathbf{x}(k_c) \\ \mathbf{y}(k_c) \end{bmatrix} \quad (31)$$

where $\mathbf{K} \in \mathbb{R}^{n \times 3n}$ is the constant gain matrix given by

$$\mathbf{K} = (\tilde{\mathbf{R}} + \tilde{\mathbf{B}}^T \mathbf{P} \tilde{\mathbf{B}})^{-1} \tilde{\mathbf{B}}^T \mathbf{P} \tilde{\mathbf{A}} \quad (32)$$

and $\mathbf{P}$ is a unique positive semidefinite solution of the discrete-time matrix algebraic Riccati equation

$$\mathbf{P} = \tilde{\mathbf{Q}} + \tilde{\mathbf{A}}^T (\mathbf{P} - \mathbf{P}\tilde{\mathbf{B}}(\tilde{\mathbf{R}} + \tilde{\mathbf{B}}^T \mathbf{P} \tilde{\mathbf{B}})^{-1} \tilde{\mathbf{B}}^T \mathbf{P}) \tilde{\mathbf{A}}. \quad (33)$$

One way to calculate $\mathbf{P}$ is by iterating backwards in time (starting from any terminal positive semidefinite condition, e.g. $\mathbf{P}(K_c) = \tilde{\mathbf{Q}}$ or $\mathbf{P}(K_c) = \mathbf{0}$) the dynamic Riccati equation of the finite-horizon case

$$\mathbf{P}(k-1) = \tilde{\mathbf{Q}} + \tilde{\mathbf{A}}^T (\mathbf{P}(k) - \mathbf{P}(k)\tilde{\mathbf{B}}(\tilde{\mathbf{R}} + \tilde{\mathbf{B}}^T \mathbf{P}(k) \tilde{\mathbf{B}})^{-1} \tilde{\mathbf{B}}^T \mathbf{P}(k)) \tilde{\mathbf{A}} \quad (34)$$

until it converges to a stationary value. This is possible only for $0 \leq \sigma < 1$ because then the system is controllable. However, if one selects $\sigma = 1$, then again gain matrix $\mathbf{K}$ can be calculated by iterating backwards in time the dynamic Riccati equation of the finite-horizon case until $\mathbf{K}$ converges to a stationary value, as the system is then not controllable, but stabilizable.

Decomposing the gain matrix $\mathbf{K} = [\mathbf{K}_1, \mathbf{K}_2]$, we have from (31)

$$\Delta\mathbf{u}(k_c) = -\mathbf{K}_1 \Delta\mathbf{x}(k_c) - \mathbf{K}_2 \mathbf{y}(k_c) \quad (35)$$

and, after some algebra, the final form of the Linear-Quadratic regulator with Integral action (LQI regulator) for the internal boundary control problem is given by the following differential form

$$\boldsymbol{\varepsilon}(k_c) = \boldsymbol{\varepsilon}(k_c-1) - \mathbf{K}_P[\mathbf{x}(k_c) - \mathbf{x}(k_c-1)] - \mathbf{K}_I\left[\tilde{\boldsymbol{\rho}}^a(k_c) - \tilde{\boldsymbol{\rho}}^b(k_c)\right] \qquad (36)$$

where $\mathbf{K}_P = \mathbf{K}_1 - \mathbf{K}_2\mathbf{H}$ and $\mathbf{K}_I = \mathbf{K}_2$ are the proportional and integral gain matrices, respectively, while $\boldsymbol{\varepsilon}(k_c) = [\varepsilon_1(k_c), \ldots, \varepsilon_n(k_c)]^T$, $\mathbf{x}(k_c) = [\tilde{\boldsymbol{\rho}}^a(k_c)^T, \tilde{\boldsymbol{\rho}}^b(k_c)^T, \boldsymbol{\gamma}(k_c)^T]^T$, $\tilde{\boldsymbol{\rho}}^a(k_c) = [\tilde{\rho}_1^a(k_c), \ldots, \tilde{\rho}_n^a(k_c)]^T$, $\tilde{\boldsymbol{\rho}}^b(k_c) = [\tilde{\rho}_1^b(k_c), \ldots, \tilde{\rho}_n^b(k_c)]^T$ and $\boldsymbol{\gamma}(k_c) = [\gamma_1(k_c), \ldots, \gamma_n(k_c)]^T$.

Some comments should be given regarding this regulator:

- No nominal values are explicitly required in the regulator.
- If the weighting matrix $\mathbf{S}$ is set equal to a zero matrix, then $\mathbf{K}_2$ becomes also a zero matrix, i.e. (35) becomes an LQ regulator without integral terms.
- It is immediately seen that, at a steady state, the last term of the regulator must equal zero, and, since $\mathbf{K}_2$ is full rank, this corresponds to equal relative densities on the two directions of every section, as specified.
- The regulator is utterly reactive, i.e. it makes use only of real-time density measurements in each section and direction. In particular, there is no need for external demand prediction, as necessary in the QP problem of Section 2.3.
- The values obtained for each one of the control variables, i.e. the sharing factors per section, must be truncated before application in order to satisfy (3). These truncated values are used as $\boldsymbol{\varepsilon}(k_c - 1)$ in (36) in the next time-step to avoid the well-known wind-up effect of regulators with integral terms. In case $\mathbf{S}$ is set equal to a zero matrix, i.e. for an LQ regulator, the differential form (36) is equivalent to (35) only if no truncation is necessary.
- Finally, (4) and (5) are applied before actual implementation of the sharing factors to ensure safe evacuation of the direction being restricted by the last control decision.

## 4. Case studies

### 4.1. Simulation set-up

The performance of the proposed feedback-based controllers is investigated using two scenarios for a bi-directional highway stretch depicted in Figure 3. These scenarios have been first considered by Malekzadeh et al. (2021) to check the performance of the optimal QP methodology proposed therein. Using the same scenarios here, allows us to compare the efficiency improvements achieved by the two LQ control methods. The two demand scenarios were carefully selected to challenge the controller more than in real traffic, as they feature very strongly rising and falling inflow values that require accordingly fast controller reaction. The two demand scenarios belong to the following respective classes (see (Malekzadeh et al., 2021) for a more in-depth discussion):

- *Uncongested scenario*, where congestion is created in one or both directions without internal boundary control (no-control case); but congestion can be utterly avoided with activation of internal boundary control. Such situations are likely to constitute the majority of real congestion cases on highways.

- *Congested scenario*, where no-control congestion can be mitigated with internal boundary control, but cannot be utterly suppressed due to strong and strongly overlapping bi-directional external demands.

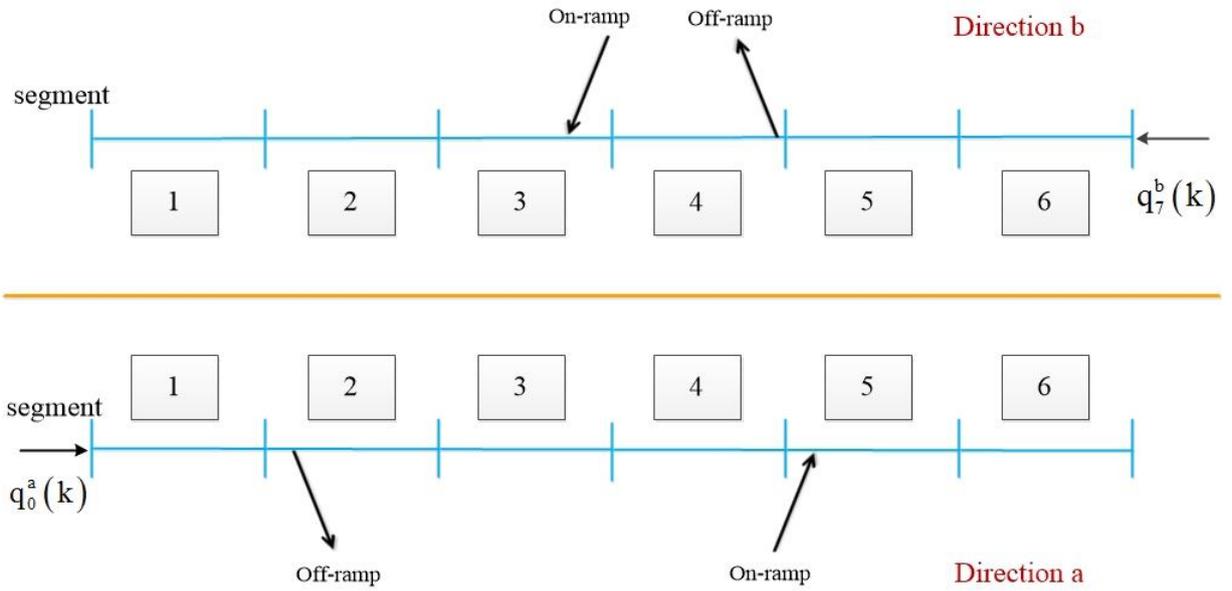

Figure 3: The considered highway stretch

The considered highway stretch has a length of 3 km and is subdivided in 6 sections of 0.5 km each. For direction $a$, there is an off-ramp in section 2 and an on-ramp in section 5, while for direction $b$, there is an off-ramp in section 4 and an on-ramp in section 3. The modelling time-step, $T$, is set to 10 s and the considered time horizon is 1 h, hence $K = 360$.

While a linearization of CTM was used for controller design, the full nonlinear CTM is used to represent the emulated ground truth in this section. The model parameters used are $v_f = 100$ km/h and $w_s = 12$ km/h; while the total cross-road capacity to be shared among the two directions is $q_{cap} = 12{,}000$ veh/h. Based on the above and the triangular FD, we can calculate $\rho_{cr} = 120$ veh/km and $\rho_{max} = 1120$ veh/km. The parameter values used to enable capacity drop, when this possibility is mentioned to be activated in the scenarios, are $\lambda_r = 0.7$ and $\lambda_d = 0.4$; when no capacity drop is activated in the model, these parameters are $\lambda_r = 1$ and $\lambda_d = 0$. The exit rates for the two off-ramps are both assumed equal to 0.1. For both scenarios, the initial density values used are $\rho^a(0) = [5.0, 5.0, 5.0, 5.0, 18.5, 29.4]$ veh/km, $\rho^b(0) = [14.4, 14.4, 14, 5.0, 5.0, 5.0]$ veh/km. For each scenario, the simulation results of the no-control case are presented first for completeness and comparison purposes, although they are already included in (Malekzadeh et al., 2021). The no-control cases are followed by the results obtained when using the LQ and LQI regulators. The efficiency improvements achieved are also presented and compared to those obtained when using the optimal control solution reported by Malekzadeh et al. (2021).

**4.2. Regulator design**

In order to apply the feedback-based LQ and LQI regulators (36) developed in Section 3, we need to calculate off-line the static gain matrix $\mathbf{K} = [\mathbf{K}_1, \mathbf{K}_2]$. The convex combination parameter is selected to be $\sigma = 0.95$. A nominal point of operation is selected for the calculation, via (37)-(39), of the matrices $\mathbf{A}$ and $\mathbf{B}$ used in the linear model (26). The controller has to be designed to perform well around the critical area of the fundamental diagram for each direction. To this end, it is assume that the nominal point for the relative densities is equal to 1. The selection of the nominal demand values is done such that the resulting mainstream flows are roughly around the capacity of the nominal system, i.e. the system with sharing factors equal to 0.5. Under these conditions, the exact nominal demand values did not have any noticeable impact on the control results. In the simulations performed, we have strong variations of the demand over time, which do not have a negative effect on the performance of the LQ and LQI controllers that is approaching the performance of optimal control. In this sense, the selected nominal values are $q_0^{aN} = q_7^{bN} = 5000$ veh/h, $r_5^{aN} = r_3^{bN} = 1000$ veh/h, $\tilde{\rho}_i^{aN} = \tilde{\rho}_i^{bN} = 1$ and $\varepsilon_i^N = 0.5$, $i = 1, 2, \ldots, 6$. From (15), we have that $\gamma_i^N = 0.5$, $i = 1, 2, \ldots, 6$. The control time-step, $T_c$, is set to 60 s, hence $K_c = 60$ and $M = 6$. Based on (30) and the $\mathbf{A}$ and $\mathbf{B}$ matrices, we can calculate the $\tilde{\mathbf{A}}$ and $\tilde{\mathbf{B}}$ matrices for the augmented linear system. The weighing matrices used in the objective function (29) are selected to be $\mathbf{Q} = [\mathbf{I}_{2n \times 2n}, \mathbf{0}_{n \times n}; \mathbf{0}_{n \times 3n}]$, $\mathbf{S} = 10^{p_1} \mathbf{I}_{n \times n}$ and $\mathbf{R} = 10^{p_2} \mathbf{I}_{n \times n}$, where the values for the parameters $p_1$ and $p_2$, appearing as exponents in the above specification of the weighing matrices $\mathbf{S}$ and $\mathbf{R}$, respectively, will be used in Sections 4.3.3 and 4.4.3 to investigate the robustness of the controller. The selection of the $\mathbf{Q}$ matrix is such that the control variables are not considered (indirectly, via the states $\Delta\boldsymbol{\gamma}$) twice in the objective function. The above specifications allow for the calculation of $\tilde{\mathbf{Q}}$ and $\tilde{\mathbf{R}}$. Then, starting from the terminal condition $\mathbf{P}(K_c) = \mathbf{0}$, we iterate backwards in time the dynamic Riccati equation of the finite-horizon case (34) until it converges to a stationary value **P** that is used in (32) to get the gain matrix **K**.

The LQI regulator (36) starts with initial control input values set equal to the nominal values, i.e. equal to 0.5 for all sections. The upper and lower bounds for the sharing factors, used to avoid utter blocking of any of the two directions, are equal for all sections $i = 1, 2, \ldots, 6$ and are given the values $\varepsilon_{i,\min} = 0.16$ and $\varepsilon_{i,\max} = 0.84$.

With these settings, the regulator is operated in a closed-loop mode, receiving in emulated real time all section density values per direction from the CTM model equations; and responding with the sharing factors, which are calculated according to (36) and are eventually truncated (if necessary) and delayed as mentioned earlier. This is repeated every $T_c = 60\,\text{s}$.

### 4.3. Uncongested scenario

#### 4.3.1. Scenario description

The mainstream and on-ramp demand flows per direction for this scenario are presented in Figure 4a. Compared to real freeways, the demand rising and falling profiles applied are quite extreme for such a short time period. It may be seen that the two directions feature respective peaks in their mainstream demands that are slightly overlapping. In addition, the on-ramp demands are constant, with the on-ramp demand in direction $a$ being higher than in direction $b$.

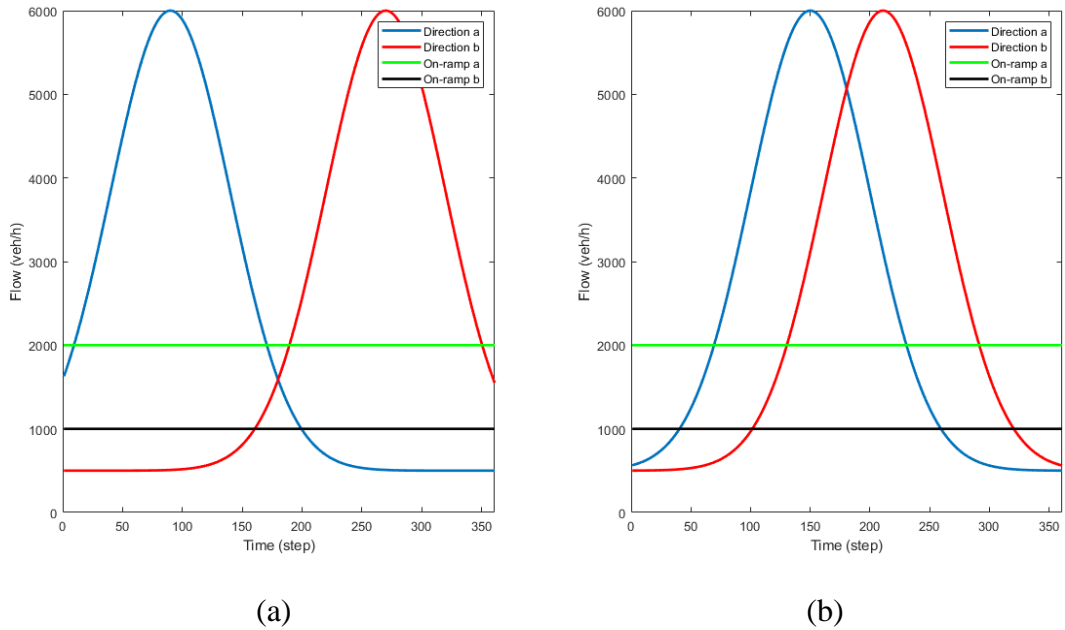

(a)                  (b)

Figure 4: Demand flows per direction and on-ramp for (a) the uncongested scenario; and (b) the congested scenario

### 4.3.2. No-control case

Using the demand flows of the uncongested scenario in the CTM equations of Section 2.2 with constant sharing factors at $\varepsilon_i = 0.5$ for all sections, we obtain the simulation results of the no-control case. Figure 5 displays the corresponding spatio-temporal evolution of the relative density defined in (12). According to the definition, relative density values lower than 1 refer to uncongested traffic; while values higher than 1 refer to congested traffic; when the relative density equals 1, and the downstream section is uncongested, we have capacity flow at the corresponding section.

Figure 5 shows that heavy congestion is created in section 5 for direction $a$ due to the strong ramp inflow, in combination with the increased mainstream demand, at around $k = 60$. The congestion propagates upstream, reaching up to section 2, and dissolves at around $k = 200$, due to the rapid decrease of the mainstream demand for this direction (Figure 4a). In direction $b$, we have also a congestion being triggered in section 3 by the increasing mainstream demand, in combination with the on-ramp flow, at around $k = 250$. Due to lower on-ramp flow, the congestion extent is smaller than in direction $a$; the congestion propagates upstream to section 5 and dissolves at around $k = 330$.

It should be noted that the results displayed in Figure 5 were obtained using the CTM equations with the capacity drop terms. The TTS value obtained for this case is reported in Table 1. When the capacity drop terms are de-activated, then the space-time extent of the created congestions is slightly reduced. The corresponding diagrams are omitted for space economy, but the resulting, lower TTS value is also reported in Table 1.

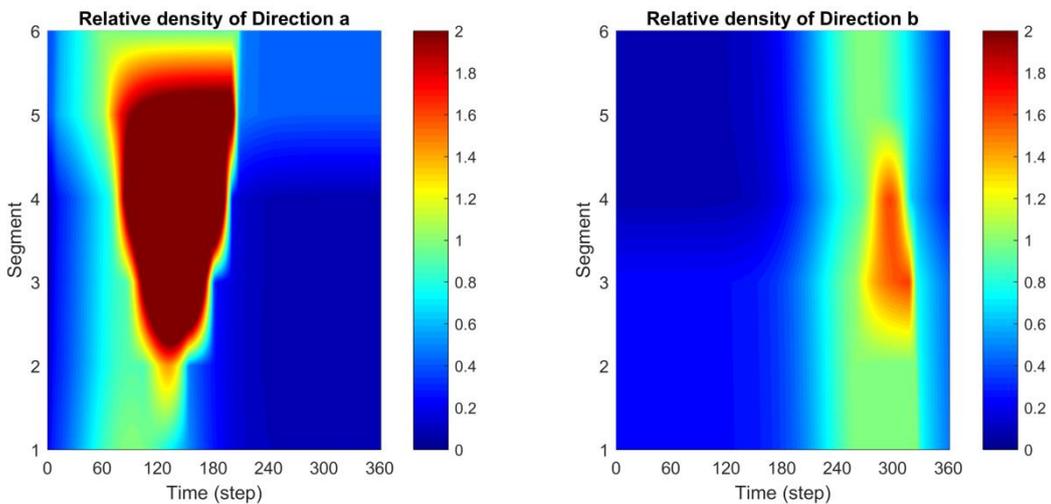

Figure 5: Uncongested scenario: Relative density for the two directions in the no-control case

Table 1: TTS (veh·h) and related improvement (%) over the no-control case for different scenarios

| Scenarios | No-control | LQI with $(p_1, p_2) = (-2.5, -3.0)$ | LQ with $(p_1, p_2) = (-\infty, -3.0)$ | QP |
|---|---|---|---|---|
| Uncongested with capacity drop | 231.9 | 164.9 (28.9) | 164.9 (28.9) | 164.9 (28.9) |
| Uncongested without capacity drop | 209.8 | 164.9 (21.4) | 164.9 (21.4) | 164.9 (21.4) |
| Congested with capacity drop | 236.0 | 170.3 (27.8) | 175.6 (25.6) | 171.0 (27.5) |
| Congested without capacity drop | 213.9 | 170.0 (20.5) | 172.7 (19.3) | 170.9 (20.1) |

#### 4.3.3. Control case

In order to investigate the robustness properties of the LQ regulators, in particular the influence of the weighting matrices in the quadratic objective (29) on control performance, we have calculated 1000 different values of the gain matrix **K**, following the procedure of Section 4.2, with randomly selected values for the parameters $p_1$ and $p_2$ within the range $[-5, 2]$. The corresponding values for the gain matrix **K**, reflecting different combinations of the weighting matrices, were then used to run a set of 1000 corresponding control experiments for the uncongested scenario. Each such experiment delivered a corresponding TTS value from the simulator, and the TTS values for all $(p_1, p_2)$ combinations are depicted, marked as black dots, in Figure 6a. The same plot includes a fitted function.

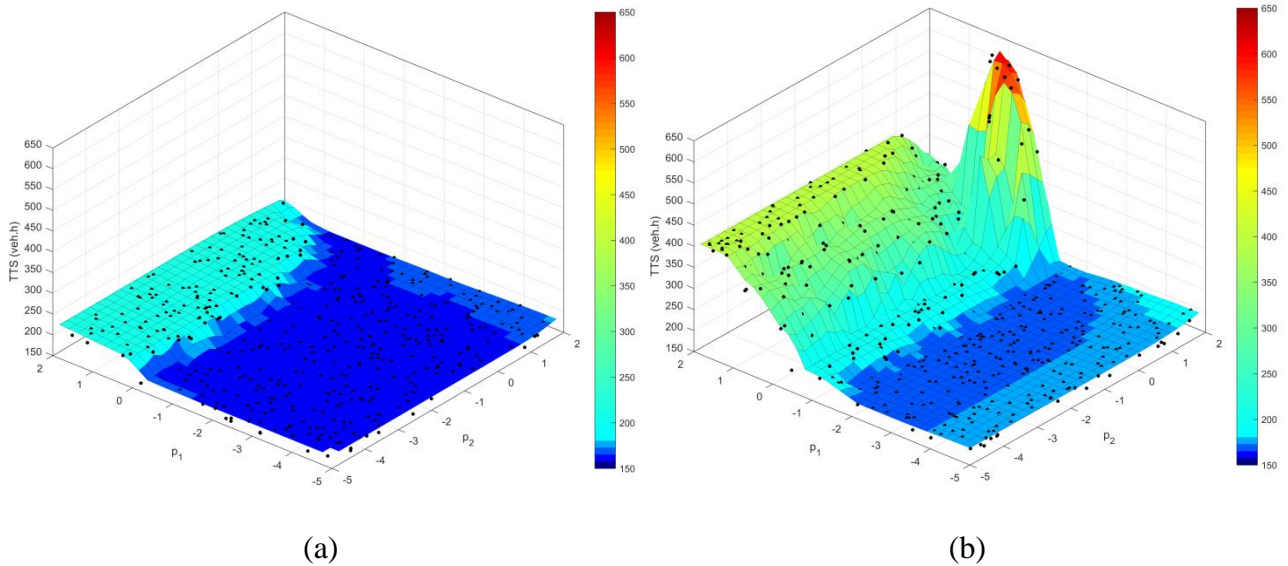

(a)          (b)

Figure 6: 3-D $(p_1, p_2)$-diagram of the TTS values for (a) the uncongested scenario; and (b) the congested scenario

Note that the 3-D diagram in Figure 6a actually refers to the LQI controller results. On the other hand, the LQ regulator is obtained for weight matrix $\mathbf{S} = \mathbf{0}$, i.e. for $p_1 \to -\infty$, in the objective function, but the displayed results for $p_1 = -5$ are virtually equivalent to those obtained with $\mathbf{S} = \mathbf{0}$. Thus, the robustness diagram for the LQ regulator without integral parts is 2-D (depending on $p_2$ only) and is obtained from the 3-D diagram of Figure 6a as the 2-D plane cut at $p_1 = -5$. This 2-D diagram is visible in Figure 6a, at the front-right edge of the 3-D diagram (for $p_1 = -5$).

Based on the fitted function, it can be safely concluded that the LQ and LQI regulators are robust with respect to variations in the values of the selected weighing matrices within a very large area, that is seen to remain flat, featuring similarly low TTS values.

In the following we have selected to use the LQI regulator with a parameter set $(p_1, p_2) = (-2.5, -3.0)$, which is within the good-performance area of Figure 6a; and the LQ regulator with the same weight $p_2$, i.e. $(p_1, p_2) = (-\infty, -3.0)$. The resulting traffic conditions for both regulators are virtually the same and under-critical anywhere anytime, i.e. congestion observed in both traffic directions in the no-control case is utterly avoided. The TTS values for both regulators are exactly the same, and the activation of the capacity drop terms has no impact, as there is no congestion. All related TTS values are given in Table 1, indicating improvements of 29 % and 21 % over the corresponding no-control cases with and without capacity drop activation, respectively. The TTS value obtained using the LQI or the LQ regulators is, in fact, also equal to the value that is achieved when applying the optimal control resulting from the QP problem formulation. Thus, despite their simple feedback character, where no model or demand predictions are involved, the LQI and LQ regulators achieve highest efficiency, even though they do not explicitly consider TTS in their objective.

Since the performance of the two regulators is equivalent, we use only the LQ regulator results for the following detailed discussion of the control-case for the uncongested scenario. The spatio-temporal evolution of the relative densities is depicted in Figure 7. Figure 8, Figure 9 and Figure 10 display more detailed information for this case. Specifically, each figure holds two columns with the results of two respective sections; for each section (column), we provide three diagrams (rows):

- The first diagram shows the two traffic densities (in veh/km), for directions *a* and *b*, and the corresponding critical densities, which are changing according to the sharing factor in the section.

- The second diagram shows the two traffic flows, for directions *a* and *b*, and the corresponding capacities, which are changing according to the sharing factor in the section. In addition, the sum of both flows is also displayed (cyan curve).
- The third diagram shows the value of the control input, i.e. the sharing factor applied, as well as the constant bounds (black curves), which may lead to possible truncation of the control input.

The displayed results confirm that densities (flows) are always lower than the respective (controlled) critical densities (capacities) in all sections and in both directions; hence traffic conditions are always and everywhere under-critical. In fact, the total-flow curve (for both directions) does not reach the total road capacity (of 12,000 veh/h) at any time anywhere. In short, congestion is utterly avoided and any occurring delays in the no-control case do not exist anymore.

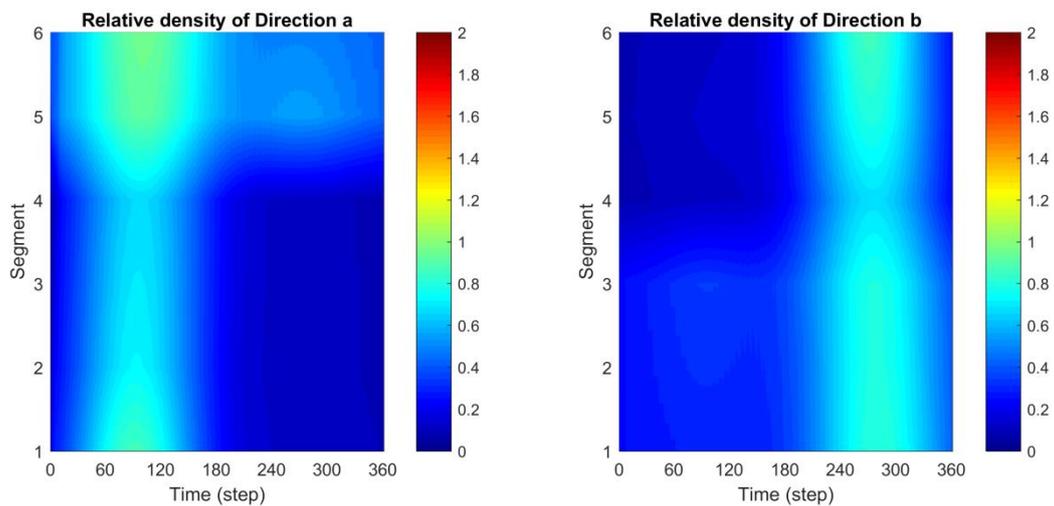

Figure 7: Uncongested scenario: Relative density for the two directions in the control case (LQ)

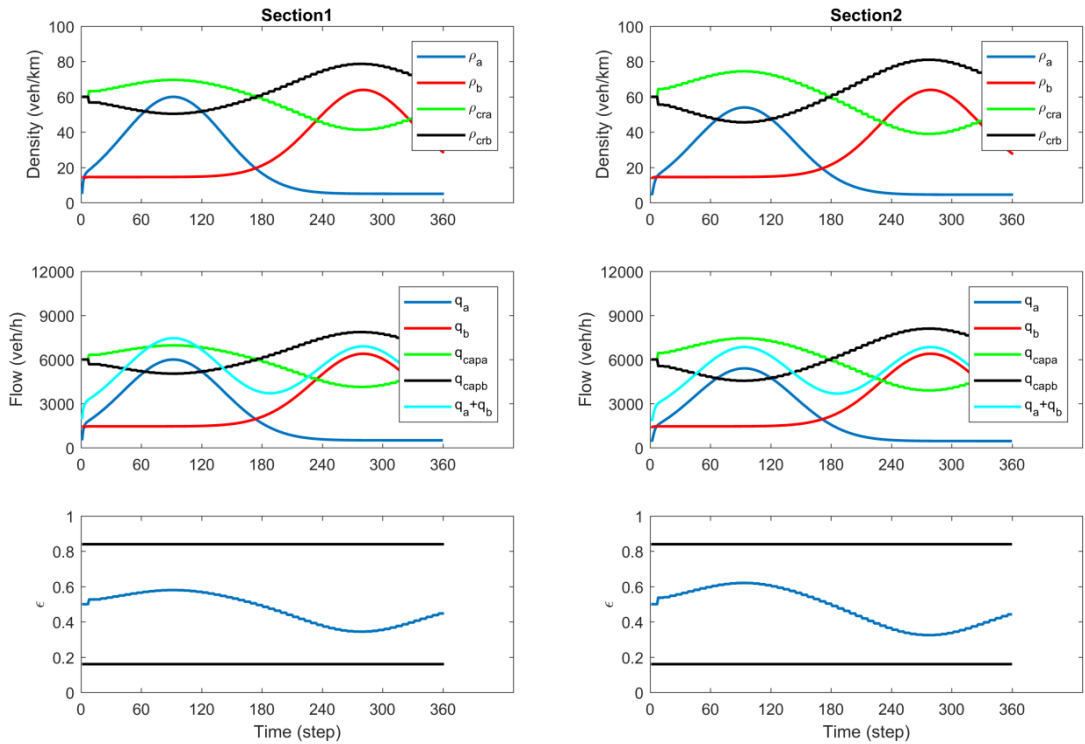

Figure 8: Uncongested scenario: Density, flow and control trajectories in the control case (LQ) (sections 1 and 2)

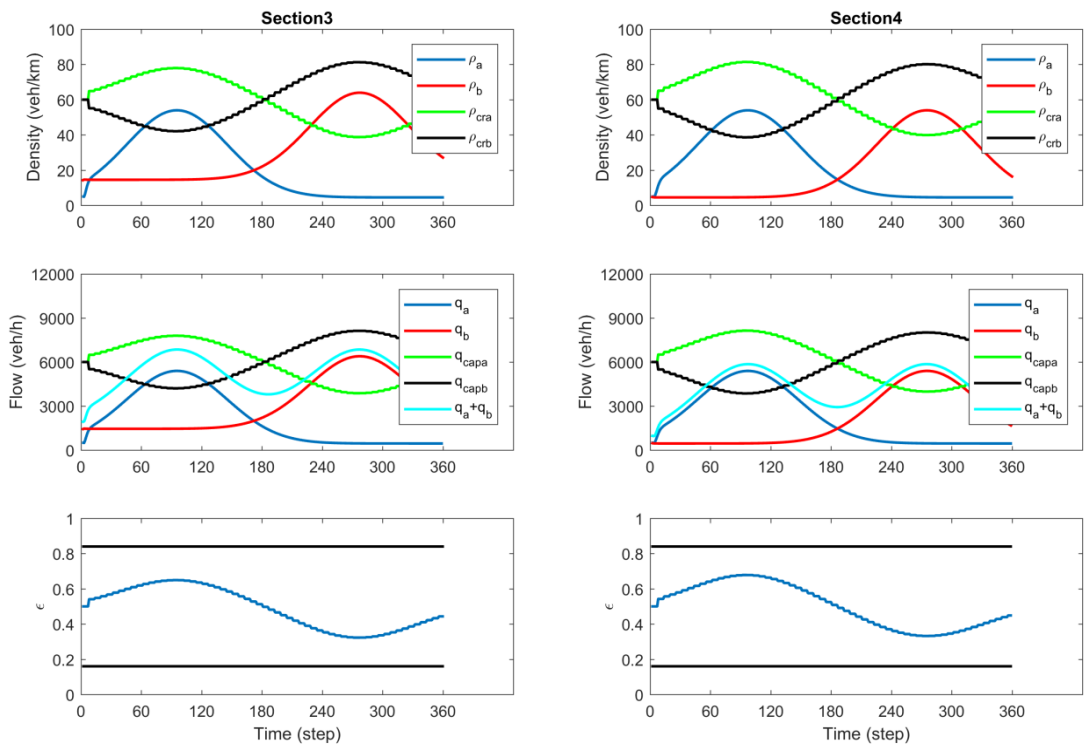

Figure 9: Uncongested scenario: Density, flow and control trajectories in the control case (LQ) (sections 3 and 4)

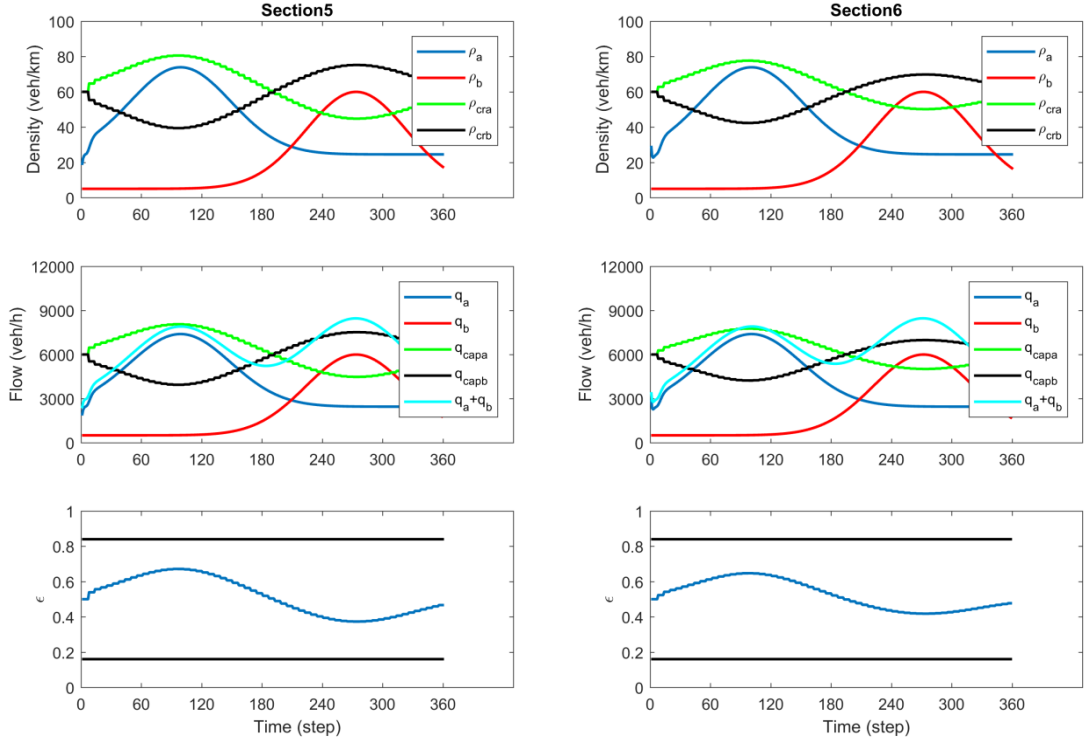

Figure 10: Uncongested scenario: Density, flow and control trajectories in the control case (LQ) (sections 5 and 6)

The sharing factor trajectories of the sections reveal that this excellent outcome is enabled via a smooth swapping of assigned capacity to the two directions, whereby more capacity is assigned to direction $a$ during the first half of the time horizon and vice-versa for the second half, in response to the traffic (density) changes caused by the changing respective demands and their peaks. It is interesting to notice that the value of the control input (sharing factor) is never saturated and, as a result, the use of the differential form (36) for the LQ controller is equivalent to the use of (35). Finally, Figure 11a displays the space-time diagram of the control input which demonstrates that it is a smooth function in space and time.

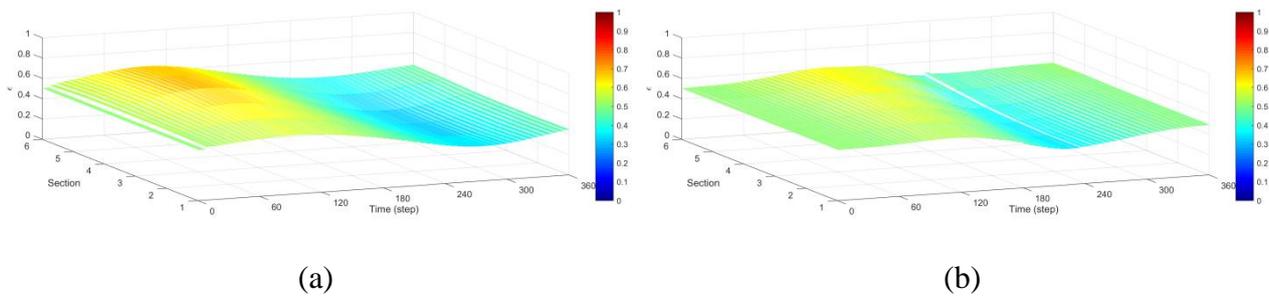

(a)  (b)

Figure 11: 3-D space-time diagram of the LQ control input for (a) the uncongested scenario; and (b) the congested scenario

### 4.4. Congested scenario

#### 4.4.1. Scenario description

The demand flows for this scenario are displayed in Figure 4b for both directions. Essentially, all external flows are similar as in the uncongested scenario, with the notable difference that the two mainstream demand profiles have been shifted closer to each other. This leads to a longer overlapping period with strong flows on both directions, which implies capacity problems even in presence of internal boundary control. The on-ramp demands are constant at the same respective levels as in the uncongested scenario.

#### 4.4.2. No-control case

Using the demand flows of the congested scenario in the CTM equations of Section 2.2 with constant internal boundary at $\varepsilon_i = 0.5$ for all sections, we obtain the simulation results of the no-control case. Figure 12 displays the corresponding spatio-temporal relative density evolution, where heavy congestion is created in section 5 for direction $a$ due to the strong ramp inflow, in combination with the increased mainstream demand, at around $k = 120$. The congestion propagates upstream, reaching up to section 2, and is dissolved at around $k = 250$, thanks to the rapid decrease of the mainstream demand for this direction (Figure 4b). In direction $b$, we have also a congestion being triggered in section 3 by the increasing mainstream demand, in combination with the on-ramp flow, at around $k = 200$. Due to lower on-ramp flow, the congestion extent is smaller than in direction $a$. The congestion propagates upstream up to section 5 and dissolves at around $k = 270$.

The results displayed in Figure 12 were obtained using the CTM equations with the capacity drop terms, and the corresponding value of TTS is reported in Table 1. When the capacity drop terms are de-activated, then the space-time extent of the created congestions is reduced. The corresponding diagrams are omitted for space economy, but the resulting, lower TTS value is also reported in Table 1. Note that, compared with the uncongested scenario, the congestion extent and TTS values are not much different, simply because the total demand per direction did not change much, while the capacity of each direction is constant in the no-control case. In other words, any observed congestion phenomena in the two directions are independent of each other and only dependent on the corresponding one-directional demand and capacity, which are both the same in the two scenarios for the no-control case.

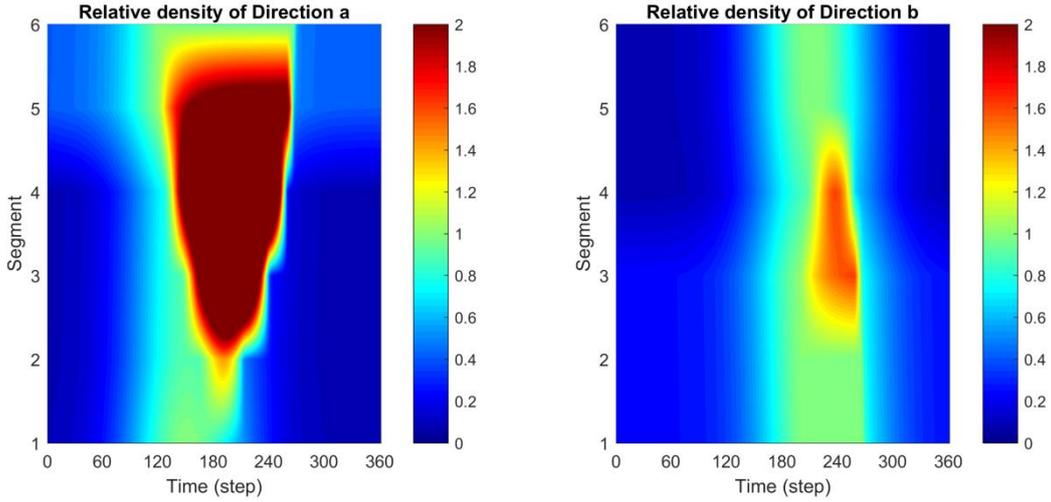

Figure 12: Congested scenario: Relative density for the two directions in the no-control case

### 4.4.3. Control case

In order to further investigate the robustness properties of the LQI and LQ regulators with respect to the objective criterion weights, we used the 1000 different values of the gain matrix **K**, calculated in Section 4.3.3, to run a set of 1000 corresponding control experiments, now for the congested scenario. Each such experiment delivered a corresponding TTS value from the simulator, and the TTS values for all $(p_1, p_2)$ combinations are depicted, marked as black dots, in Figure 6b. Again, the same plot includes a fitted function. Recall that the 2-D diagram for the LQ regulator appears in Figure 6b at the front-right edge of the 3-D diagram (for $p_1 = -5$).

Comparing the two diagrams of Figure 6, we observe that:

- The good-performance area has slightly shrunk in the congested scenario compared to the uncongested one, but remains very large, extending over several orders of magnitude of the weights $p_1$ and $p_2$.

- The good-performance area for the congested scenario is a subset of its counterpart of the uncongested scenario. This implies that the same weights may be used for high-performance regulation in either case.

- The best control performance of the LQI regulator is slightly better compared to the LQ regulator.

- In case of inappropriate weight selection, the worst performance of the LQ regulator (obtained for $p_2 \to \infty$) is equivalent to no control. In contrast, inappropriate weight selection for the LQI regulator may lead to results much worse than no control.

Further scenario investigations with stronger congestion confirmed that both regulators perform excellently (i.e. deliver TTS values that are not more than 5 % higher than the optimal ones) under all scenarios, for $p_1 < -2$ and $p_2 < 0$.

In fact, for the present demand scenario, the LQI regulator features similarly low TTS values, virtually equal to the one achieved when applying the optimal control resulting from the QP problem formulation, within a broad area of $(p_1, p_2)$ combinations, marked blue in Figure 6b; thus, the selection of any pair within this area, e.g. $(p_1, p_2) = (-2.5, -3.0)$, leads to optimal results. On the other hand, for $p_1 < -4$ the matrix $\mathbf{S} = 10^{p_1} \mathbf{I}_{n \times n}$ is approaching the zero matrix, i.e. the LQI regulator is approaching the LQ regulator, and we observe that for $p_2 < 0$ (and $p_1 = -5$) the LQ regulator is producing close to optimal results (see also Table 1). It must be emphasized that this excellent outcome of the feedback regulators is achieved without assuming a perfect model and perfect demand prediction, as for the optimal QP problem.

Activation of the capacity drop terms has only a minor impact, as there is no heavy congestion in the control results. The TTS value obtained using the LQI regulator (see Table 1) is even slightly lower than the value resulting from the optimal control of the QP problem of Section 2.3. This may sound paradoxical, but is explained by the fact that the optimal QP problem solution takes into account, apart from TTS, also other operational sub-objectives in (11), which may slightly affect the resulting TTS value upon minimization.

The results reported in the following figures were obtained with activation of the capacity drop terms. Figure 13 displays the spatio-temporal evolution of the relative densities in the control case, utilizing the LQI regulator with $(p_1, p_2) = (-2.5, -3.0)$, while Figure 14 displays the same output when utilizing the LQ regulator with $(p_1, p_2) = (-\infty, -3.0)$. It can be observed that in both cases a short-lasting congestion exists in section 6 for direction $b$, while the main difference is the congestion created in section 5 for direction $a$ when the LQ regulator is used, which is absent in the LQI regulator results.

Figure 15, Figure 16 and Figure 17 display more detailed information for the case of the LQ regulator, as for the uncongested scenario. The displayed results indicate that in direction $a$ congestion appears in section 5, starting at time $k = 150$ and lasting up to $k = 225$. In direction $b$, congestion is observed in section 6, starting at time $k = 175$ and lasting up to $k = 225$. It is important to emphasize that the total-flow curve (for both directions) reaches and remains close to the total road capacity (of 12,000 veh/h) at sections 5 and 6 during this peak period. In fact, sections 5 and 6 form a bottleneck for this scenario. As explained in (Malekzadeh et al., 2021), bottlenecks

in the internal boundary control context, under an efficient control strategy, refer to both traffic directions simultaneously, which is contrast to independent uni-directional bottlenecks appearing in conventional traffic control. Thus, full exploitation of the road capacity (both directions) is indeed enabled at the bottleneck by the LQ regulator for the duration of the critical period, so as to minimize congestion and delays.

The sharing factor trajectories are smooth in this scenario as well (see Figure 11b), achieving the efficient assignment of the necessary share of capacity where and when needed. No saturation is observed for the control input (sharing factors) at any section.

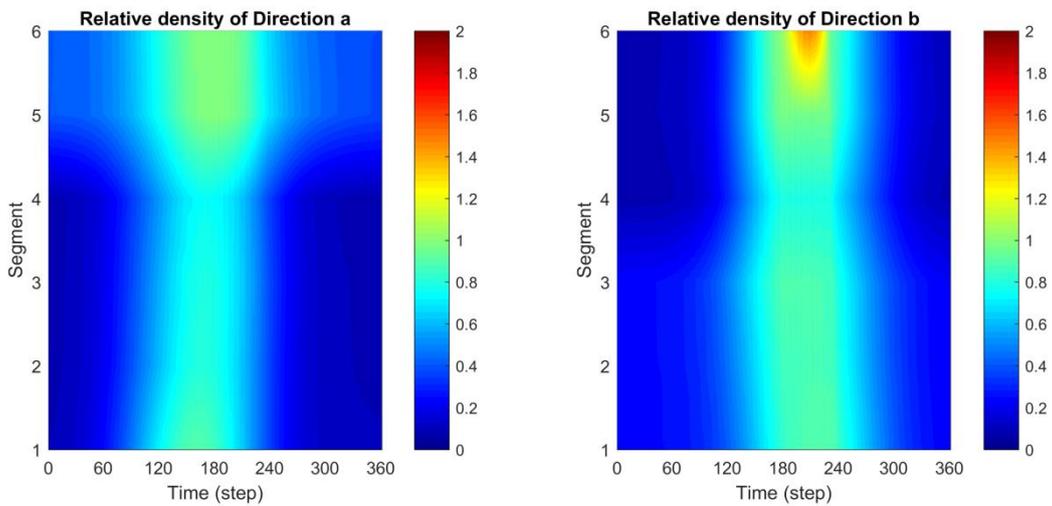

Figure 13: Congested scenario: Relative density for the two directions in the control case (LQI)

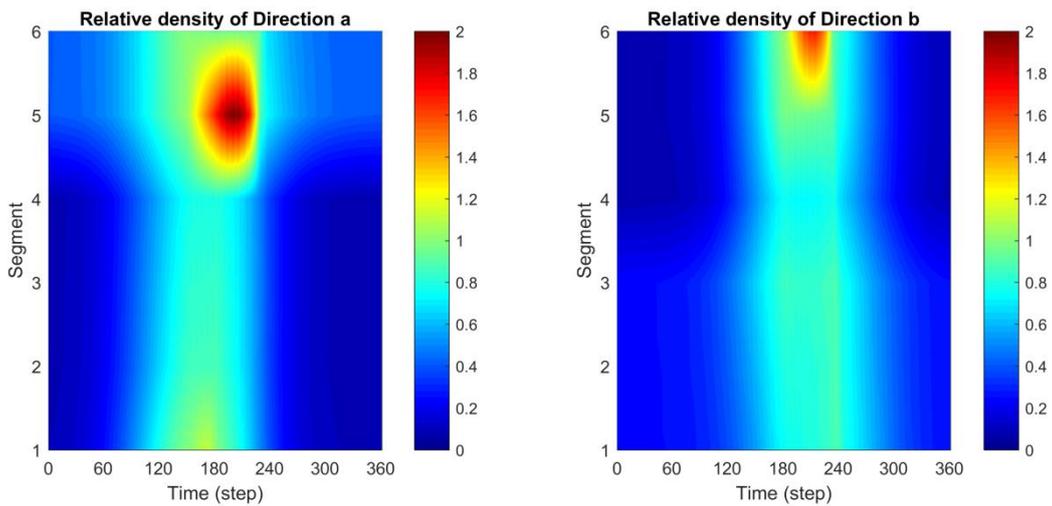

Figure 14: Congested scenario: Relative density for the two directions in the control case (LQ)

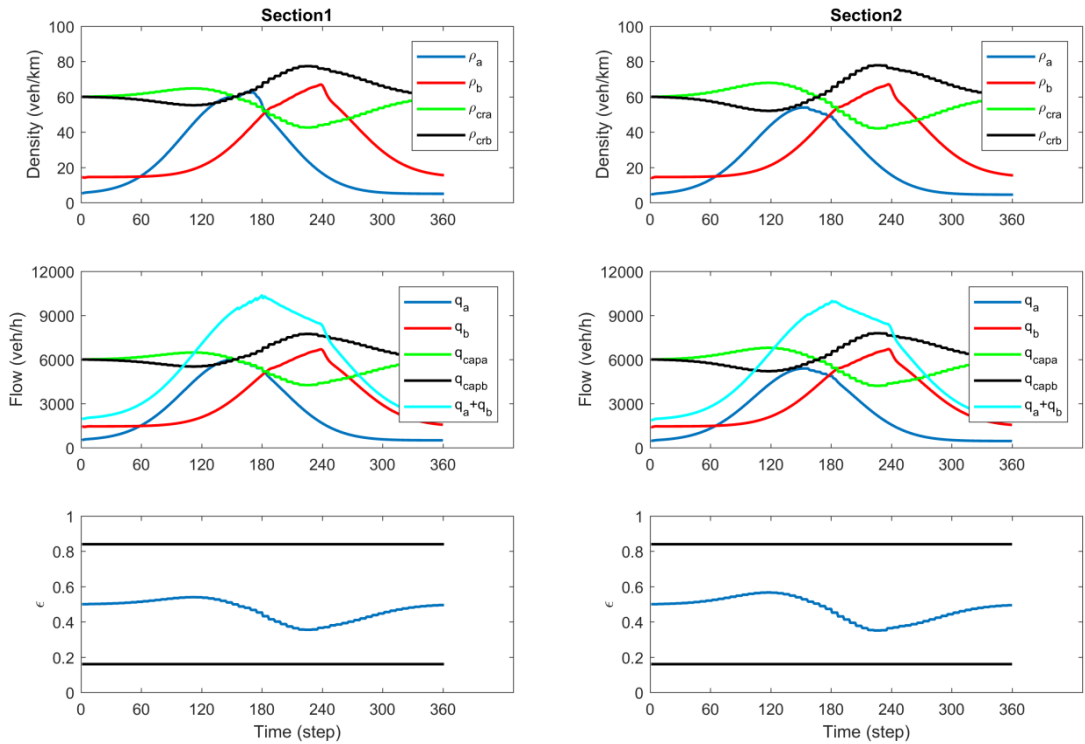

Figure 15: Congested scenario: Density, flow and control trajectories in the control case (LQ) (sections 1 and 2)

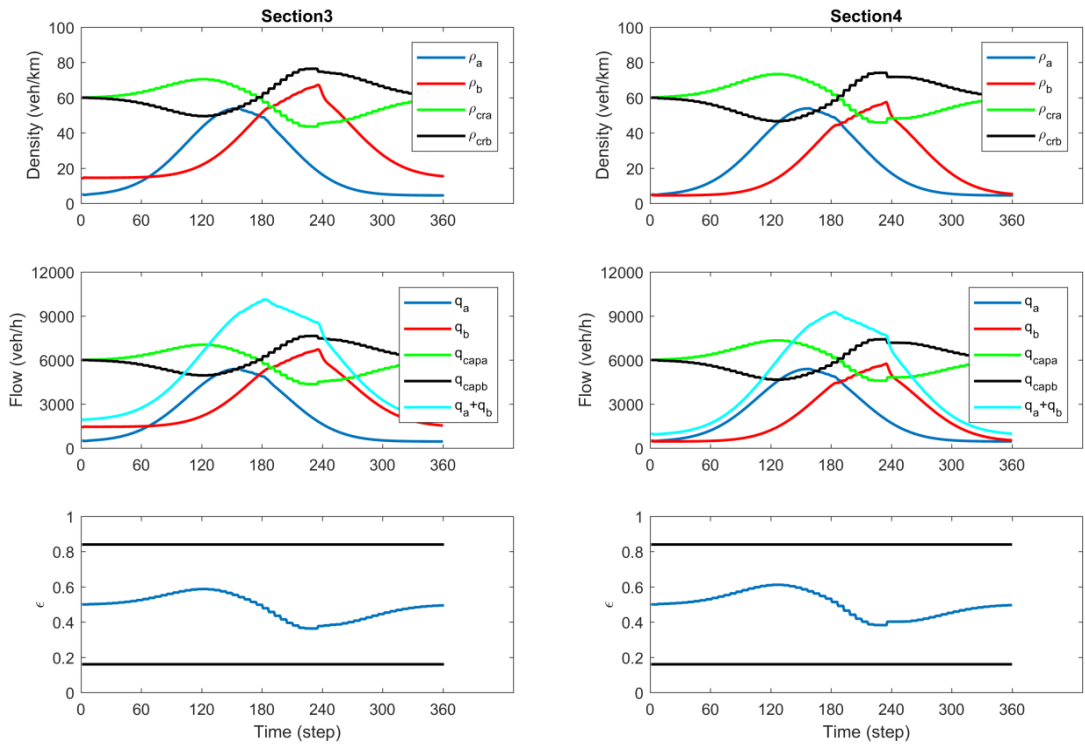

Figure 16: Congested scenario: Density, flow and control trajectories in the control case (LQ) (sections 3 and 4)

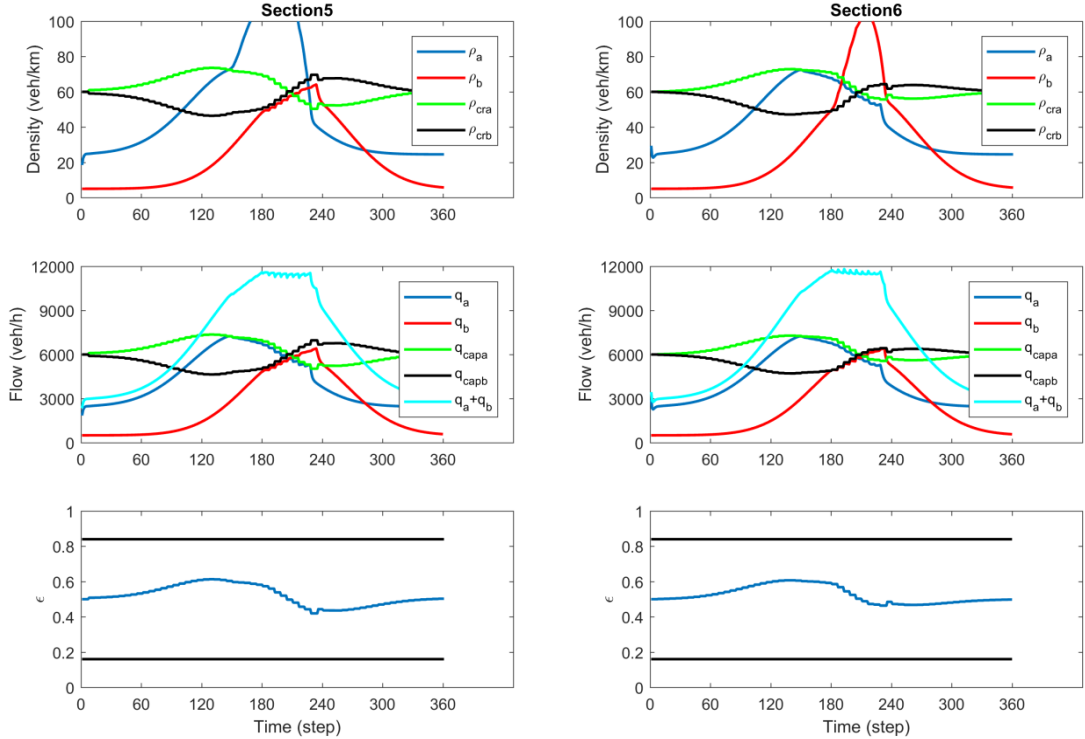

Figure 17: Congested scenario: Density, flow and control trajectories in the control case (LQ) (sections 5 and 6)

### 4.5. Uncongested scenario with late control activation

In the scenarios considered already, the LQ regulator is able to either completely avoid congestion (when possible) or minimize the extent of congestion in space and time (when congestion is unavoidable). In the considered scenarios, the regulator is applied for the whole simulation period, starting from under-critical initial values of the traffic state. In order to test and demonstrate that the regulator is successful even in cases where congestion has already occurred, we simulate the system for the uncongested scenario without applying any control action for an initial period of time, which includes the appearance of congestion due to lack of control, as in the no-control case. Then, at $k = 72$, the LQ regulator is activated and remains active till the end of the simulation.

The resulting spatio-temporal evolution of the relative densities is depicted in Figure 18. As expected, up to $k = 72$, traffic conditions are exactly the same with those of the no-control case described in Section 4.3.2 and depicted in Figure 5. The onset of congestion takes place at around $k = 60$ in section 5 of direction $a$ and propagates upstream to section 3. At $k = 72$, thanks to the activation of the controller, the congestion propagation is mitigated, and its extent starts soon after to decrease rapidly.

Figure 19, Figure 20 and Figure 21 display more detailed information for this case. The actions of the controller lead to an increase of the assigned capacity for direction $a$, immediately after the activation of the controller, in reaction to the accumulated demand in the congested area. Congestion is dissolved at around $k=170$, much earlier compared to the case of no control activation. Despite this action, direction $b$ remains uncongested, as, in the second half of the simulation, more capacity is assigned to this direction to accommodate its own rising demand. The TTS value achieved with the late activation of the LQ regulator is 180.6 veh·h, which is a 22% improvement compared to the case of no control activation reported in Table 1.

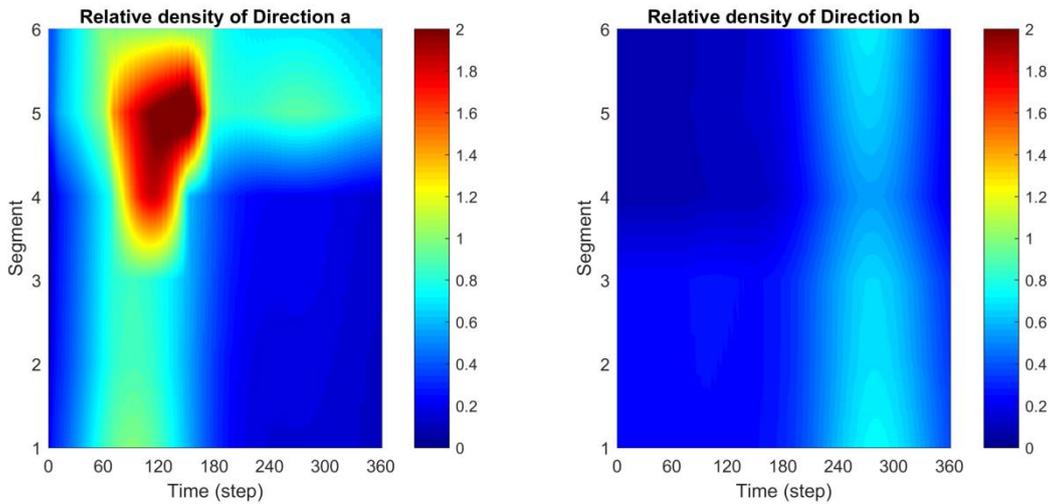

Figure 18: Uncongested scenario with late control (LQ) activation: Relative density for the two directions

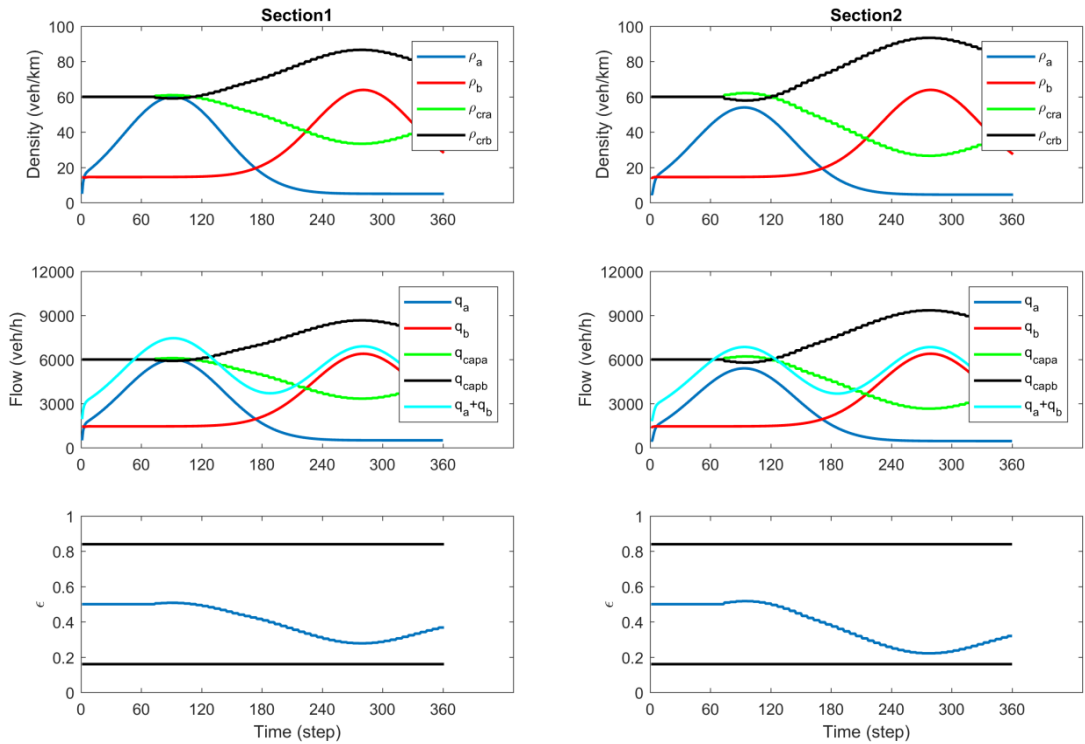

Figure 19: Uncongested scenario with late control (LQ) activation: Density, flow and control trajectories (sections 1 and 2)

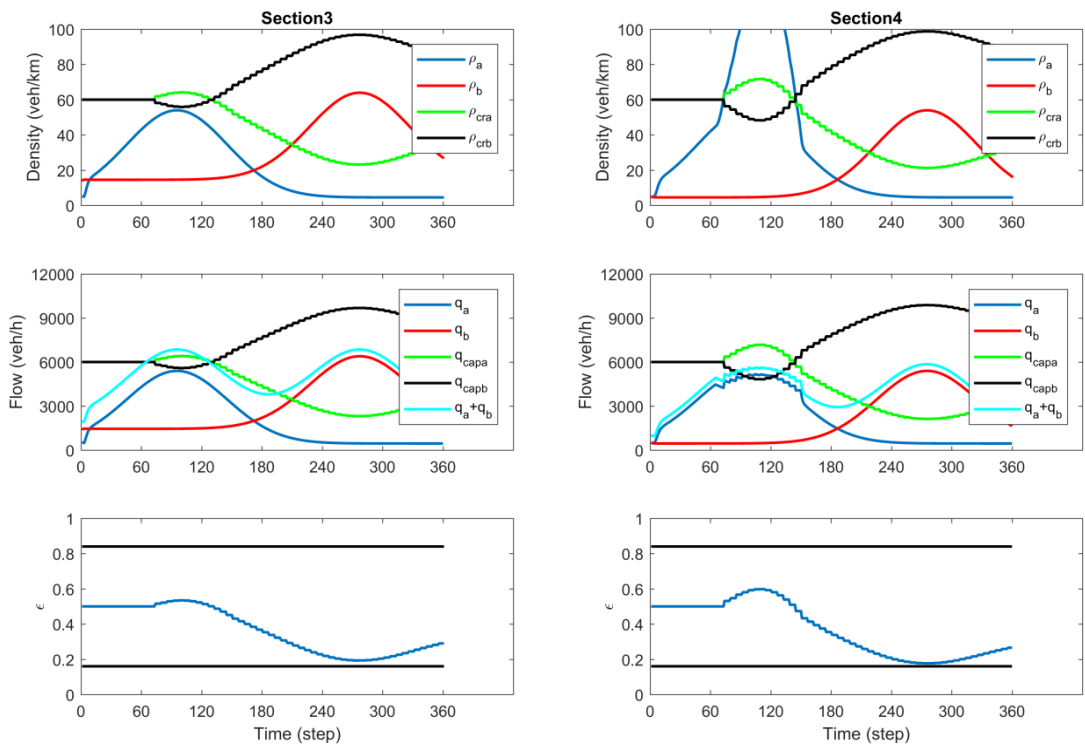

Figure 20: Uncongested scenario with late control (LQ) activation: Density, flow and control trajectories (sections 3 and 4)

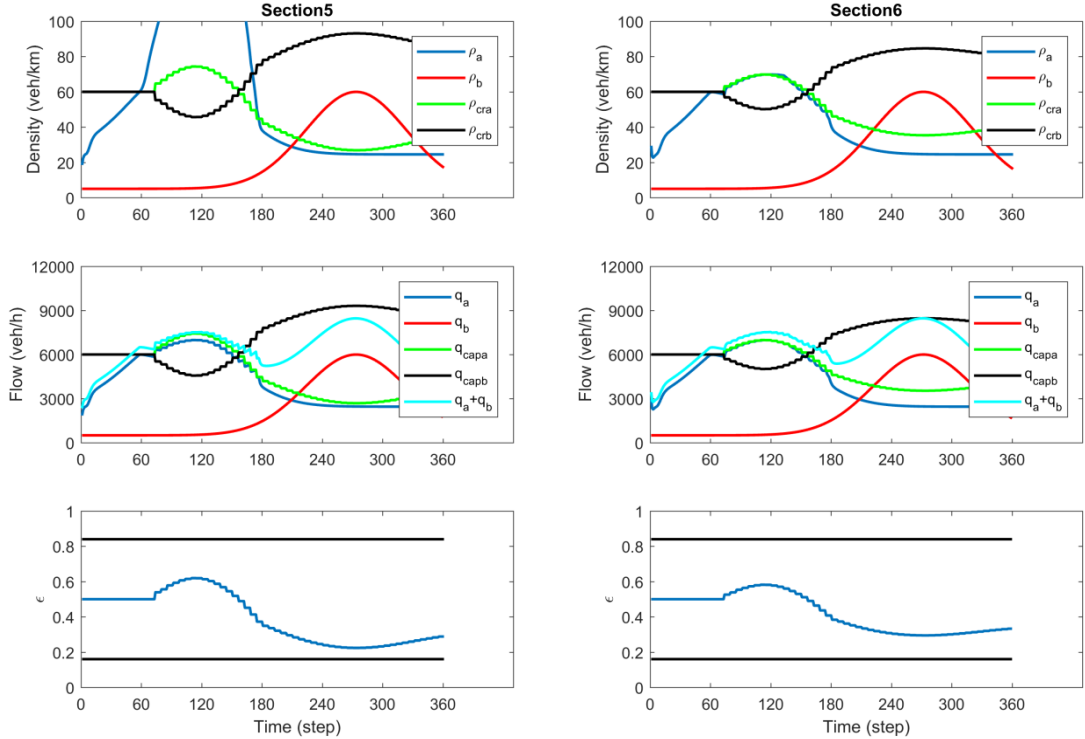

Figure 21: Uncongested scenario with late control (LQ) activation: Density, flow and control trajectories (sections 5 and 6)

## 5. Conclusions and future work

The concept of internal boundary control, introduced by Malekzadeh et al. (2021), has been revisited in this study by use of a different control approach. The well-known CTM, appropriately adjusted to introduce the effect of the sharing factors, has been utilized for the design of feedback-based LQ and LQI regulators for the internal boundary control problem. Using these controllers, the total road width and capacity are shared in each section in real-time among the two directions of the road in response to the prevailing traffic conditions, without the need for model and demand predictions. The regulators were found to behave adequately for a broad range of weight values used in their quadratic objective and for different traffic scenarios, in which congestion may be avoidable or unavoidable.

The LQ and LQI regulators are easy to design and implement (feedback-based) and robust to disturbances (no need to predict the arriving demands). Simulation investigations demonstrate, for carefully selected scenarios, that the regulators (with appropriate selection of the weighting matrices) are equally efficient as an open-loop optimal control solution (with perfect model and demand predictions) developed for the same problem by Malekzadeh et al. (2021) using a convex QP problem formulation.

Ongoing work considers microscopic simulation studies, with vehicles moving in a lane-free mode based on appropriate CAV movement strategies. Specific physical traffic parameters that appear in the linearized CTM model and are necessary for the calculation of the gain matrix for the controller, such as free speed, critical density, flow capacity, will be calibrated based on data produced by the microscopic simulator. Given the feedback character of the LQ regulators, possible moderate inaccuracies in the specification of these parameters (which would reflect in corresponding minor differences in the gain matrix elements) would not have a significant impact on the control results.

Furthermore, we also plan to consider concept testing for real large-scale highway infrastructure scenarios.

**Acknowledgments**

The research leading to these results has received funding from the European Research Council under the European Union's Horizon 2020 Research and Innovation programme/ ERC Grant Agreement n. [833915], project TrafficFluid.

**Appendix A**

This Appendix presents the derivation of the partial derivatives necessary in (23)-(25). For direction $a$, the partial derivatives necessary for (23) are as follows:

$$\left.\frac{\partial f_i^a}{\partial \tilde{\rho}_i^a}\right|_N = 1 + \frac{T}{L_i \rho_{cr}}\left[-(1-\sigma)v_f \gamma_i \rho_{cr}\right]\bigg|_N, \quad i=1,2,\ldots,n$$

$$\left.\frac{\partial f_i^a}{\partial \tilde{\rho}_{i-1}^a}\right|_N = \frac{T}{L_i \rho_{cr}}\left[(1-\beta_i^a)\frac{(1-\sigma)v_f \gamma_{i-1}\rho_{cr}}{\gamma_i}\right]\bigg|_N, \quad i=2,3,\ldots,n$$

$$\left.\frac{\partial f_i^a}{\partial \varepsilon_i}\right|_N = \frac{T}{L_i \rho_{cr}}\left[-\frac{\sigma q_{cap}}{\gamma_i}\right]\bigg|_N, \quad i=1,2,\ldots,n$$

$$\left.\frac{\partial f_i^a}{\partial \varepsilon_{i-1}}\right|_N = \frac{T}{L_i \rho_{cr}}\left[(1-\beta_i^a)\frac{\sigma q_{cap}}{\gamma_i}\right]\bigg|_N, \quad i=2,3,\ldots,n$$

$$\left.\frac{\partial f_i^a}{\partial \gamma_i}\right|_N = \frac{T}{L_i \rho_{cr}}\left[-(1-\beta_i^a)\frac{\sigma\varepsilon_{i-1}q_{cap}+(1-\sigma)v_f\tilde{\rho}_{i-1}^a\gamma_{i-1}\rho_{cr}}{\gamma_i^2}+\frac{\sigma\varepsilon q_{cap}}{\gamma_i^2}-\frac{d_i^a}{\gamma_i^2}\right]\bigg|_N, \quad i=2,3,\ldots,n$$

$$\left.\frac{\partial f_i^a}{\partial \gamma_{i-1}}\right|_N = \frac{T}{L_i \rho_{cr}}\left[(1-\beta_i^a)\frac{(1-\sigma)v_f \tilde{\rho}_{i-1}^a \rho_{cr}}{\gamma_i}\right]\bigg|_N, \quad i=2,3,\ldots,n$$

$$\left.\frac{\partial f_1^a}{\partial \gamma_1}\right|_N = \frac{T}{L_1 \rho_{cr}}\left[-\frac{d_1^a}{\gamma_1^2}+\frac{\sigma\varepsilon_1 q_{cap}}{\gamma_1^2}\right]\bigg|_N \quad (37)$$

Similarly for direction $b$, the partial derivatives necessary for (24) are as follows:

$$\left.\frac{\partial f_i^b}{\partial \tilde{\rho}_i^b}\right|_N = 1 + \frac{T}{L_i \rho_{cr}}\left[-(1-\sigma)v_f \rho_{cr}\right]_N, \; i=1,2,\ldots,n$$

$$\left.\frac{\partial f_i^b}{\partial \tilde{\rho}_{i+1}^b}\right|_N = \frac{T}{L_i \rho_{cr}}\left[(1-\beta_i^b)\frac{(1-\sigma)v_f(1-\gamma_{i+1})\rho_{cr}}{1-\gamma_i}\right]_N, \; i=1,2,\ldots,n-1$$

$$\left.\frac{\partial f_i^b}{\partial \varepsilon_i}\right|_N = \frac{T}{L_i \rho_{cr}}\left[\frac{\sigma q_{cap}}{1-\gamma_i}\right]_N, \; i=1,2,\ldots,n$$

$$\left.\frac{\partial f_i^b}{\partial \varepsilon_{i+1}}\right|_N = \frac{T}{L_i \rho_{cr}}\left[-(1-\beta_i^b)\frac{\sigma q_{cap}}{1-\gamma_i}\right]_N, \; i=1,2,\ldots,n-1$$

$$\left.\frac{\partial f_i^b}{\partial \gamma_i}\right|_N = \frac{T}{L_i \rho_{cr}}\left[(1-\beta_i^b)\frac{\sigma(1-\varepsilon_{i+1})q_{cap}+(1-\sigma)v_f \tilde{\rho}_{i+1}^b(1-\gamma_{i+1})\rho_{cr}}{(1-\gamma_i)^2} - \frac{\sigma(1-\varepsilon_i)q_{cap}}{(1-\gamma_i)^2} + \frac{d_i^b}{(1-\gamma_i)^2}\right]_N,$$
$$i=1,2,\ldots,n-1$$

$$\left.\frac{\partial f_i^b}{\partial \gamma_{i+1}}\right|_N = \frac{T}{L_i \rho_{cr}}\left[-(1-\beta_i^b)\frac{(1-\sigma)v_f \tilde{\rho}_{i+1}^b \rho_{cr}}{1-\gamma_i}\right]_N, \; i=1,2,\ldots,n-1$$

$$\left.\frac{\partial f_n^b}{\partial \gamma_n}\right|_N = \frac{T}{L_n \rho_{cr}}\left[\frac{d_n^b}{(1-\gamma_n)^2} - \frac{\sigma(1-\varepsilon_n)q_{cap}}{(1-\gamma_n)^2}\right]_N \qquad (38)$$

Moreover, we have simply

$$\left.\frac{\partial g_i}{\partial \varepsilon_i}\right|_N = 1, \; i=1,2,\ldots,n. \qquad (39)$$